\documentclass[11pt, letterpaper]{article}

\usepackage{graphicx}
\usepackage{epstopdf}
\usepackage{amsmath,amssymb,amsfonts}
\usepackage{setspace}
\usepackage{epstopdf}
\usepackage{color}
\usepackage{xcolor}
\usepackage[letterpaper,left=1in,right=1in,top=1in,bottom=1in]{geometry}
\usepackage{multirow}
\usepackage{longtable}
\usepackage{rotating}
\usepackage{mathrsfs}
\usepackage{algorithm}
\usepackage{algorithmicx}
\usepackage{algpseudocode}
\usepackage{booktabs}
\usepackage{ntheorem}
\usepackage{arydshln}
\usepackage{subfig} 
\usepackage[round]{natbib}
\usepackage[colorlinks=true,allcolors=blue]{hyperref}
\usepackage{enumerate}
\usepackage{bm}
\usepackage{makecell}
\usepackage{lineno}  

\title{\Large \bf Economic zone data-enabled predictive control for connected open water systems}

\author{
\centerline{\normalsize Xiaoqiao Chen$^{a,b,c}$, Xuewen Zhang$^{c}$, Minghao Han$^{a}$, Adrian Wing-Keung Law$^{d,}$\thanks{Corresponding author: A. W.-K. Law. Tel: (+65) 6516 2273. Email: cewklaw@nus.edu.sg.}, 
Xunyuan Yin$^{a,c,}$\thanks{Corresponding author: X. Yin. Tel: (+65) 6316 8746. Email: xunyuan.yin@ntu.edu.sg.}
}
\vspace{5mm}\\
\centerline{\small $^{a}$Nanyang Environment and Water Research Institute, Nanyang Technological University, }\\
\centerline{\small 1 CleanTech Loop, 637141, Singapore}\\
\centerline{\small $^{b}$Interdisciplinary Graduate Programme, Nanyang Technological University,}\\
\centerline{\small 61 Nanyang Drive, 637460, Singapore}\\
\centerline{\small $^{c}$School of Chemistry, Chemical Engineering and Biotechnology, Nanyang Technological University,}\\
\centerline{\small 62 Nanyang Drive, 637459, Singapore}\\
\centerline{\small $^{d}$Department of Civil and Environmental Engineering, National University of Singapore,}\\
\centerline{\small 1 Engineering Drive 2, 117576, Singapore}
}

\allowdisplaybreaks

\begin{document}

\date{}
\maketitle
\setstretch{1.5}

\begin{abstract}
The real-time operation of open water systems is essential for ensuring operational safety, satisfying operational requirements, and optimizing energy usage.
However, existing rule-based control strategies rely heavily on human experience, while model-based approaches depend on accurate hydrodynamic models, which limits their applicability to water systems with complex dynamics and uncertain disturbances.
In this work, we develop a fully data-driven, zone-based control framework with adaptive control target zone selection for safe and energy-efficient operation of connected open water systems.
Specifically,  we propose a mixed-integer economic zone data-enabled predictive control (DeePC) approach that aims to maintain the water levels of the branches within the desired water-level zone while reducing real-time operational energy consumption. 
The DeePC-based approach enables direct use of input-output data for predictive control, eliminating the need for explicit dynamic modeling. 
To handle multiple control objectives with different priorities, we employ lexicographic optimization and reformulate the traditional DeePC cost function to incorporate zone tracking and energy consumption minimization objectives. 
Additionally, Bayesian optimization is utilized to determine the control target zone, which enables an effective trade-off between zone tracking and energy consumption in the presence of external disturbances. Comprehensive simulations and comparative analyses demonstrate the effectiveness of the proposed method. 
The proposed method maintains water levels within the desired water-level zone for 97.04\% of the operating time, with an average energy consumption of 33.5 kWh per 0.5 hour. Compared to baseline methods, the proposed approach reduces the frequency of zone violations by 96.95\% relative to economic zone DeePC without Bayesian optimization, and lowers energy consumption by 44.08\% relative to economic set-point tracking DeePC. As compared to rule-based control, the proposed method lowers zone-violation frequency by 74.96\% and the average energy consumption by 22.44\%.

\end{abstract}

\noindent{\bf Keywords:} data-enabled predictive control, energy consumption minimization, model predictive control, connected open water systems, water level regulation, zone tracking.

\section{Introduction}
Connected open water systems play a critical role in the urban water management cycle~\citep{castelletti2023Model}. They provide important functions such as flood mitigation, ecosystem protection, and agricultural irrigation support~\citep{horvath2022Potential,becker2024Optimization}. Achieving these objectives requires effective regulation of water levels through hydraulic structures, including pumps, weirs, and sluice gates. While preventing flooding and environmental damage are often the first priority, economic cost should also be carefully considered in control system design given the high energy consumption of pumping stations~\citep{vanderheijden2022Multimarket}. Connected open water systems are usually large-scale nonlinear systems which consist of multiple branches and hydraulic structures. The complex system dynamics make it difficult to regulate water levels and minimize economic cost with traditional control methods~\citep{castelletti2023Model}. Additionally, connected open water systems are subject to time-varying external disturbances caused by fluctuating meteorological conditions~\citep{maestre2012Distributed} and potential tidal effect at the downstream end~\citep{vanderheijden2022Multimarket}, which further pose challenges in achieving safe and energy-efficient operation. The above-mentioned factors highlight the necessity of developing advanced control methods for connected open water systems.

Model predictive control (MPC), which is an optimization-based advanced control strategy that predicts future system behaviors and determines optimal control actions subject to system constraints~\citep{morari1999Model, rawlings2017Model,christofides2013Distributed}, has been widely used for regulating the operation of water systems~\citep{castelletti2023Model, becker2024Optimization, vanderheijden2022Multimarket, maestre2012Distributed, vanoverloop2014Model, kong2023Predictive, vanoverloop2008Multiple}. 
\cite{kong2023Predictive} applied MPC to an open water system with pumping stations to regulate the water levels and reduce operating energy. In~\cite{vanoverloop2008Multiple}, multiple model predictive control was proposed for a drainage canal system to address uncertain inflows, which minimizes the risk of damage by combining the results of multiple MPC controllers. 
In practice, regulating water levels to an exact set-point is energy-intensive and unrealistic in the presence of significant disturbances and uncertainties; instead, maintaining water levels within zones is generally sufficient~\citep{horvath2022Potential}, for example, for keeping water levels within safety bounds~\citep{horvath2022Potential}, preserving adequate flood storage capacity~\citep{breckpot2013Flood}, or facilitating coordination of downstream operations~\citep{negenborn2009Distributed}. Accordingly, in~\cite{horvath2022Potential}, multi-objective first-principles model-based MPC was applied to a polder water system to maintain water levels of the branches within a predefined zone and reduce economic cost.
The objective of maintaining water levels within a predefined zone aligns naturally with the concept of zone model predictive control (zone MPC)~\citep{gonzalez2009Stable,gonzalez2009Robust,ferramosca2010MPC}, a variant of MPC that aims to keep system variables within a specified target range rather than tracking a fixed set-point~\citep{liu2019Modelpredictive,liu2018Economic}. Compared to set-point tracking, this zone-tracking formulation offers greater flexibility in control action selection and can enhance robustness to disturbances and uncertainties~\citep{liu2019Modelpredictive}, making it particularly suitable for operating connected open water systems where external disturbances and uncertainties are prevalent.  

As discussed above, pumping stations within connected open water systems can consume a significant amount of electrical energy~\citep{horvath2022Potential, vanderheijden2022Multimarket}. Therefore, in addition to maintaining water levels within the desired water-level zone, it is also important to focus on minimizing the energy consumption of the pumps during real-time system operations.
These two control objectives, including water-level zone tracking and energy consumption minimization, result in a complex optimal system operation problem involving objectives with different priority levels. This problem can be effectively addressed using lexicographic optimization~\citep{isermann1982Linear,rentmeesters1996Theory}, a multi-objective optimization approach that prioritizes objectives by importance and optimizes them sequentially, without compromising the previously optimized objectives~\citep{rentmeesters1996Theory}. Lexicographic optimization has been integrated with MPC for multi-objective optimal control of a polder water system in~\cite{horvath2022Potential}, a chemical process in~\cite{jv2023Lexicographic}, and a polymerization reactor in~\cite{anilkumar2016Lexicographic}.

However, model-based MPC methods, such as~\cite{horvath2022Potential,kong2023Predictive,vanoverloop2008Multiple}, rely on high-fidelity first-principles dynamic models to achieve accurate prediction and good control performance. Developing such models is often time-consuming and resource-intensive~\citep{rivas-perez2014Mathematical}, and their applicability to other systems is limited when the system configurations differ substantially.
In~\cite{putri2024Datadriven, zeng2025Physicsinformed, balla2022Learningbased}, data-driven MPC methods were proposed for the control of water systems, which alleviate the dependence on first-principles modeling by building dynamic models from data.
Specifically, \cite{putri2024Datadriven} developed a data-driven model for a water distribution system through system identification and proposed a data-based MPC method to control the system water levels. \cite{zeng2025Physicsinformed} incorporated a physics-informed loss into the training of an autoencoder-based Koopman model, and developed Koopman-based MPC for an open water system to track target water levels. In~\cite{balla2022Learningbased}, a data-driven MPC framework was proposed to control combined wastewater networks under varying weather conditions.
Nevertheless, these approaches still involve an explicit modeling stage, during which the system dynamics are identified or approximated prior to controller design. Moreover, the methods in \cite{putri2024Datadriven} and \cite{zeng2025Physicsinformed} only focus on tracking fixed water-level targets; the method in \cite{balla2022Learningbased} allows bounded fluctuations, but does not consider energy consumption, which is an aspect equivalently important for real-time operation of open water systems.

Model-free reinforcement learning (RL) methods have also been proposed for water-level control and energy minimization in pumping stations.
\cite{ren2021Enabling} proposed an efficient deep reinforcement learning algorithm which incorporates domain knowledge to manage water delivery tasks in a canal system. \cite{gan2024Research} used deep deterministic policy gradient (DDPG) algorithm to minimize energy consumption for pumping stations.
However, the RL training process requires extensive interactions with the underlying system, which can lead to undesirable fluctuations in water levels~\citep{sutton2018Reinforcement}.

An alternative and promising framework is data-enabled predictive control (DeePC), which formulates predictive control directly from system data without requiring first-principles knowledge or explicit system identification~\citep{berberich2021Datadriven,coulson2019Dataenabled}. In recent years, DeePC has gained increasing attention, because it solely relies on input-output data to solve constrained optimal control problems~\citep{coulson2019Dataenabled}. DeePC has been successfully applied across diverse domains, including water distribution system~\citep{perelman2025Data}, power system~\citep{huang2022Decentralized}, automated vehicle~\citep{li2024Physicsaugmented}, and chemical process~\citep{zhang2025Deep,yan2025Economic}. However, the application of DeePC in connected open water systems remains largely unexplored, and extending it to such systems is nontrivial. Conventional DeePC formulations~\citep{berberich2021Datadriven,coulson2019Dataenabled} were mainly designed for set-point tracking tasks, which are not well aligned with the operational objectives of open water systems. In particular, these methods cannot be directly used for zone tracking or energy consumption minimization, and are unable to address the disjoint input constraints of the pumps.

Building on the above literature review, there is a lack of data-driven control strategies that can simultaneously achieve water-level zone tracking and energy efficiency while addressing practical constraints in connected open water systems. In this context, zone tracking refers to maintaining the system water levels within the desired zone, which is a predefined water-level range that reflects operational requirements. Extending DeePC for zone tracking and energy consumption optimization objectives can help narrow the existing gap in data-driven control for open water systems.
Meanwhile, this also introduces a new challenge: due to external disturbances and system uncertainties, directly tracking the desired water-level zone may result in frequent zone violations. 
Therefore, a control target zone used by the DeePC-based controller, typically narrower than the desired water-level zone, need to be carefully designed as a tuning parameter to ensure satisfactory zone-tracking performance and energy efficiency under complex dynamics. To address this issue, we incorporate a Bayesian optimization-based mechanism to automatically determine the optimal control target zone, which enhances the robustness and adaptability of the proposed method for connected open water systems.

\begin{figure}[t!]
    \centering
    \includegraphics[width=1.0\textwidth]{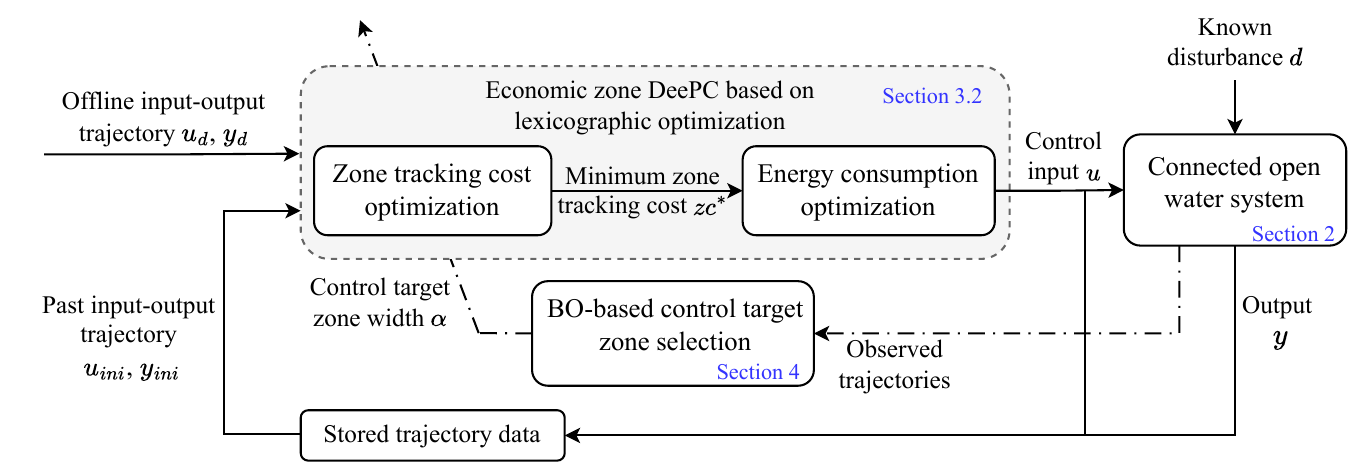}
    \caption{An illustrative diagram of the key components of the proposed method.}
    \label{fig:method_diagram}
\end{figure}

In this work, we focus on a connected open water system similar to that studied in~\cite{horvath2022Potential}. A mixed-integer economic zone DeePC framework is developed by integrating DeePC with the concepts of zone MPC and economic MPC to achieve water-level zone regulation and energy optimization using only system input-output data. 
Lexicographic optimization is exploited to formulate two optimization problems of different priority levels: water-level zone tracking and energy consumption minimization. These optimization problems are solved sequentially at each sampling instant, such that energy consumption is minimized under optimal water-level zone tracking performance.
Furthermore, Bayesian optimization is employed to identify the optimal control target zone. The effectiveness of the proposed framework is demonstrated through extensive simulations and comparative analyses against baseline control strategies.
An illustrative diagram that summarizes the key components of the proposed method is shown in Figure~\ref{fig:method_diagram}.

The novelty of this study lies in the development of a data-driven control framework that simultaneously achieves water-level zone regulation and energy optimization without requiring explicit process modeling. The main contributions of this work are summarized as follows:
\begin{itemize}
  \item We propose a mixed-integer economic zone DeePC method, which maintains the water levels within the desired water-level zone while minimizing the system energy consumption.
  \item We employ Bayesian optimization to determine the optimal control target zone in the formulated controller, which balances water-level zone tracking and operational energy efficiency in the presence of external disturbances.
  \item We conduct comprehensive simulation and comparative studies to demonstrate the effectiveness and superior performance of the proposed method.
\end{itemize}

\section{System description and problem formulation}
\begin{figure}[t!]
    \centering
    \includegraphics[width=1.0\textwidth]{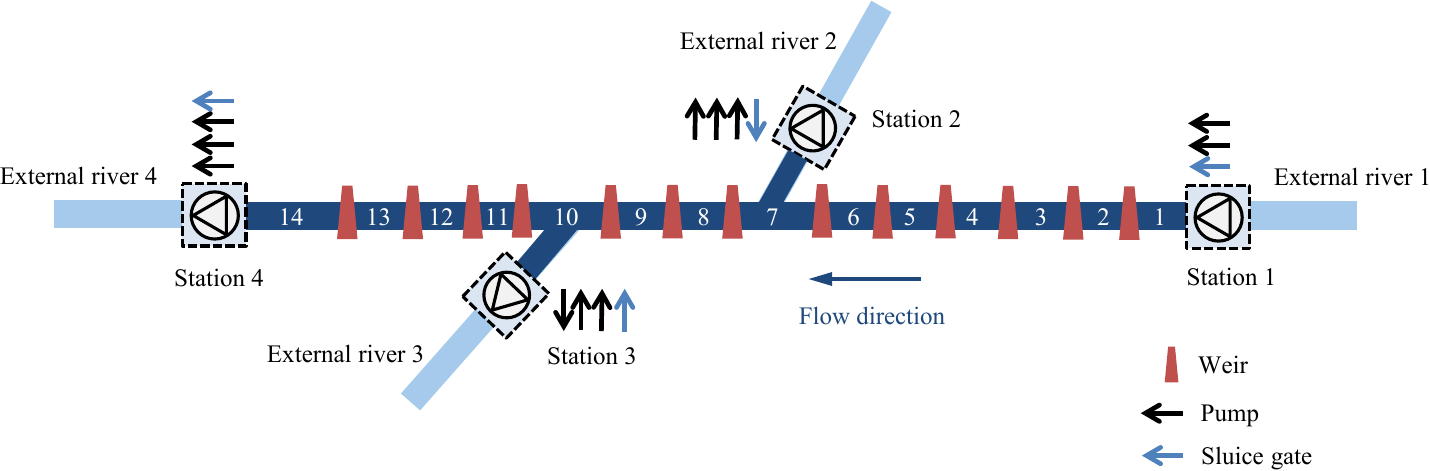}
    \caption{A schematic diagram of the connected open water system, adapted from \cite{horvath2022Potential}. The dark blue areas represent the controlled branches, and the light blue areas indicate the external rivers.
    The number of black arrows at each station corresponds to the number of pumps, with the direction of each arrow indicating the flow direction for the corresponding pump. The blue arrow indicates the permitted flow direction through the sluice gate.}
    \label{fig:Schematic}
\end{figure}

In this work, we consider a connected open water system, of which the configuration is adapted from~\cite{horvath2022Potential}. An illustrative schematic of this system is shown in Figure~\ref{fig:Schematic}. The entire system consists of multiple branches connected by hydraulic structures, including weirs, gates and pumps. Specifically, this system comprises 14 branches interconnected by 13 weirs. Additionally, four of these branches are connected to external rivers through stations, each equipped with multiple pumps and a sluice gate. The pumps and gate regulate flow across the station based on the water level difference between the two sides. The detailed configurations of the stations are provided in Table~\ref{tab:pumping_station}. The overall structure of the considered system is identical to that in~\cite{horvath2022Potential}, yet the modeling of pumps and sluice gates is different from~\cite{horvath2022Potential}. Some of the key parameters are adopted from \cite{horvath2019Closedloop}. 

{
  \renewcommand{\arraystretch}{1.2}
  \begin{table}[!t]
    \centering
    \caption{Configuration of the four stations.}
    \label{tab:pumping_station}
    \begin{tabular}{cccc}
      \toprule
      Station ID & Location (branch ID) & Pump configuration & Permitted gate flow direction \\ \midrule
      1 & 1 & 2 inflow pumps & inflow \\ 
      2 & 7 & 3 outflow pumps & inflow \\ 
      3 & 10 & 1 inflow, 2 outflow pumps & outflow \\ 
      4 & 14 & 3 outflow pumps & outflow \\ 
      \bottomrule
    \end{tabular}
  \end{table}
}

\subsection{Connected open water system}

The fluctuations of the water level within the controlled branches are represented through the conservation of mass. Specifically, each branch is modeled with the mass balance equation, expressed as~\citep{horvath2022Potential}:
\begin{equation}
  A \frac{dh}{dt} = Q_{in} - Q_{out}
\end{equation}
where $A$ is the backwater area (in m$^{2}$); $h$ is average water level of the branch (in m); $Q_{in}$ and $Q_{out}$ are the inflow and outflow discharges to the branch (in m$^{3}$/s), respectively. The discharges include the weir, pump, and gate flows, which are described in the subsequent sections, as well as external disturbance inflow to each branch.

It is worth noting that the hydrodynamics of open water systems are commonly described by the Saint--Venant (S--V) equations~\citep{litrico2009Modeling}. In this study, however, a simplified mass-balance-based simulator that retains only the continuity equation~\citep{horvath2022Potential} is adopted to evaluate the proposed data-driven control framework without relying on full hydrodynamic models. 

\subsection{Weir modeling} 
\begin{figure}[!t]
  \centering
  \subfloat[Schematic diagram of the free flow weir.]{\label{fig:weir_flow} \includegraphics[width=0.23\textwidth]{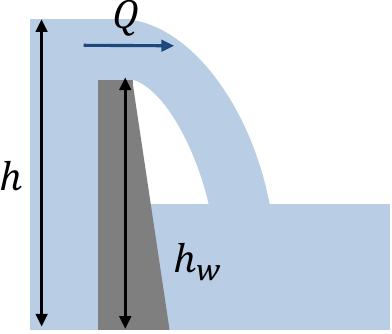}} \hspace{2cm}
  \subfloat[Schematic diagram of the inflow gate.]{\label{fig:gate_flow} \includegraphics[width=0.26\textwidth]{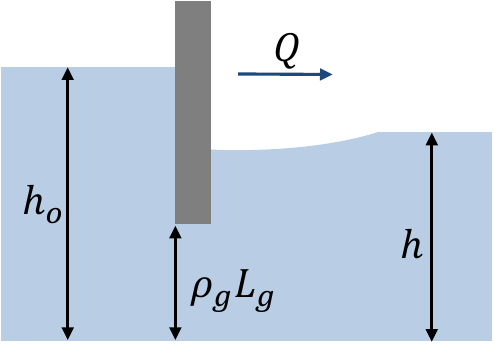}}
  \caption{Schematic diagrams of weir and gate flows.}
  \label{fig:weir_gate_flow}
\end{figure}

Weirs are used to regulate the water levels of the branches, while allowing excess water to overflow. In this connected open water system, only unsubmerged weirs are considered, where the discharge is solely determined by water head above the weir height; the lower-water-level side does not affect the flow~\citep{litrico2009Modeling}. An illustration of the weir structure and flow is shown in Figure~\ref{fig:weir_gate_flow}\subref{fig:weir_flow}. The weir discharge can be described as~\citep{horvath2022Potential}:
\begin{equation}
  Q_w = \frac{2}{3} C_{dw} w_{w} \sqrt{2g} (h-h_w)^{3/2}
\end{equation}
where $Q_w$ is the weir discharge (in m$^{3}$/s); $C_{dw}$ is the weir discharge coefficient; $w_w$ is the weir crest width (in m); $g$ is the gravitational acceleration (in m/s$^{2}$); $h$ is the water level of the higher-water-level side (in m); $h_w$ is the weir height (in m). To ensure satisfaction of the free-flow condition for the weirs, $h_w$ should lie between the water levels of the two connected branches, as described below:
\begin{equation}
  \min(h^i, h^{i+1}) \le h_{w}^{i} \le \max(h^i, h^{i+1})
\end{equation}
where $h_{w}^{i}$ is the height of the $i$th weir, for $i=1,2, \ldots ,13$; $h^i$, $h^{i+1}$ are the water levels on the two sides of the $i$th weir, corresponding to the $i$th branch and the $(i+1)$th branch, respectively, for $i=1,2, \ldots ,13$.

\subsection{Sluice gate modeling}
A sluice gate regulates the water flow by adjusting the gate opening ratio, allowing gravity-driven discharge between the system and the external environment. We consider that all the sluice gates operate under submerged flow conditions~\citep{litrico2009Modeling}. The discharge through a sluice gate is modeled as~\citep{litrico2009Modeling}:
\begin{equation}
  Q_g = C_{dg}w_g \rho_{g} L_g \sqrt{2gH_{g}}
\end{equation}
where $Q_g$ is the discharge through the gate (in m$^{3}$/s); $C_{dg}$ is the gate discharge coefficient; $w_g$ is the width of the gate (in m); $L_g$ is the maximum gate opening (in m); $H_g$ is the absolute water level difference across the gate (in m), that is, $|h-h_{o}|$, where $h$ is the water level of the branch (in m), and $h_o$ is the water level of the external river (in m). $\rho_{g} \in [0,1]$ is the gate opening ratio, which serves as one of the control inputs to the system. In this study, gate operation is constrained in a manner similar to a check valve, that is, the gate is allowed to open only when the water levels satisfy the following conditions:
\begin{equation}
  \begin{cases} 
    h_o-h\ge 0, & \text{if inflow gate (water flows into the branch)} \\
    h-h_o\ge 0 , & \text{if outflow gate (water flows out of the branch)}
  \end{cases}
\end{equation}
Otherwise, the gate input $\rho_{g}$ is set to zero. A schematic diagram of the inflow gate is shown in Figure \ref{fig:weir_gate_flow}\subref{fig:gate_flow}. For the outflow gate, $h$ is higher than $h_o$, resulting in flow in the opposite direction.

\subsection{Pump modeling\label{chap:pump_model}}
Pumps transfer water from the suction side to the discharge side. The pump discharge is determined by the shaft speed and the total head. At two operating conditions, denoted by $a$ and $b$, the performance of each pump adheres to the affinity laws~\citep{potter2011mechanics}:
\begin{equation}
  \label{eqn:pump_aff}
    \begin{aligned}
        &\frac{Q_{p,a}}{Q_{p,b}}=\frac{N_{p,a}}{N_{p,b}},\\
        &\frac{H_{p,a}}{H_{p,b}}=\left( \frac{N_{p,a}}{N_{p,b}} \right)^2
    \end{aligned}
\end{equation}
where $Q_p$ is the pump discharge (in m$^{3}$/s); $H_p$ is the total head of the pump (in m); $N_p$ is pump shaft speed (in rpm). The shaft speed of each pump also serves as one of the control inputs to the system. Given the head-discharge (H-Q) curve at nominal speed, the discharge at other operating points can be determined using \eqref{eqn:pump_aff}.

\begin{figure}[t!]
    \centering
    \subfloat[Characteristic curves and feasible operating region for each pump. Each color-coded line corresponds to one shaft speed.]{\includegraphics[width=0.435\textwidth]{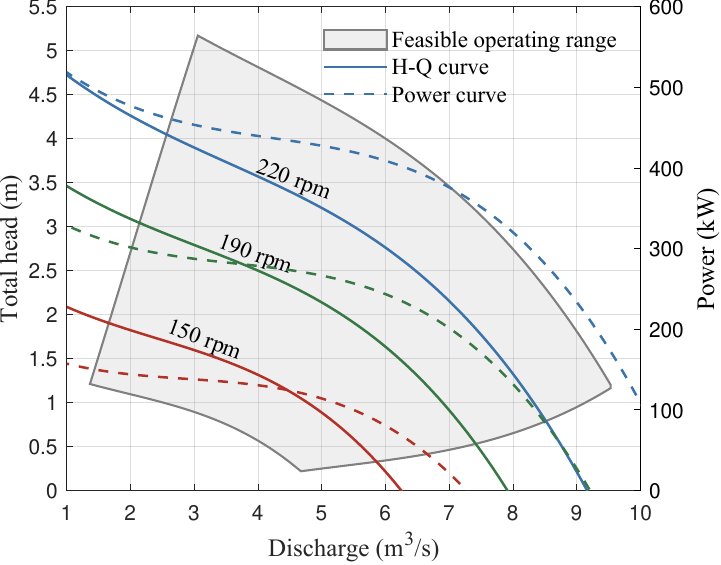} \label{fig:pump_bound}} 
    \hspace{15pt}
    \subfloat[Schematic illustrating the determination of the operating point for each pump.]{\includegraphics[width=0.4\textwidth]{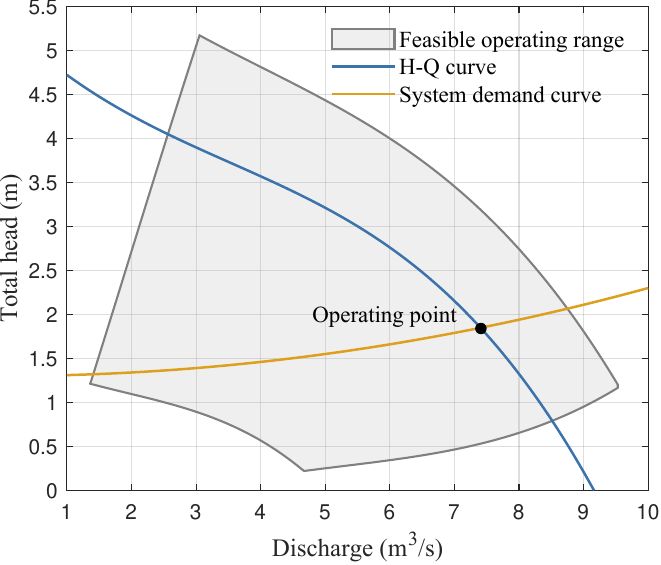} \label{fig:pump_op_point}}
    \caption{Characteristic curves, system demand curve, and feasible operating region of each pump~\citep{potter2011mechanics}.}
    \label{fig:pump_chara}
\end{figure}

The energy consumption of the open water system is primarily due to pump operations \citep{horvath2022Potential}. The power consumption of the pump can be expressed as \citep{ulanicki2008Modeling}:
\begin{equation}
  \label{eqn:pump_energy}
  \begin{aligned}
    P^{n}(Q_p) &= a_1 Q_p^3 + a_2 Q_p^2 + a_3 Q_p + a_4\\
    P_{p} &= \bar{N_{p}}^{3}P^{n}(Q_p/\bar{N_{p}}) 
           = a_1 Q_p^3 + a_2 \bar{N_{p}} Q_p^2 + a_3 \bar{N_{p}}^2 Q_p + a_4 \bar{N_{p}}^3
  \end{aligned}
\end{equation}
where $P_{p}$ is the power consumption of the pump (in kW); $\bar{N_p}=\text{current speed} / \text{nominal speed}$ is the normalized shaft speed; $P^{n}(Q_p)$ is the power characteristic curve of the pump running at nominal speed; $a_1, a_2, a_3, a_4$ are the corresponding polynomial coefficients which can be estimated using manufacturer data. The H-Q curves and power curves at different shaft speeds are presented in Figure~\ref{fig:pump_chara}\subref{fig:pump_bound}.

The feasible operating region of each pump is constrained by several factors, including minimum discharge requirements, power limitations, and the need to avoid cavitation~\citep{horvath2019Convex}. Additionally, each pump can be shut off completely, resulting in zero flow rate, shaft speed, and power consumption. 
In Figure~\ref{fig:pump_chara}\subref{fig:pump_bound}, the gray zone represents the feasible operating region for an active pump. The pump shutdown condition corresponds to the y-axis in Figure~\ref{fig:pump_chara}\subref{fig:pump_bound}, where the discharge is zero, but the total head may remain non-zero.

During operation, the pump head $H_{p}$ needs to match the hydraulic demand of the system.
To transfer water across the corresponding station, pumps are connected to pipelines that connect the branches of the station to the external rivers. For the pipeline section containing the pump, the system demand curve can be described as~\citep{potter2011mechanics}:
\begin{equation}
  H_d = H_s + \left( f_{D} \frac{L_p}{D} + \sum K \right) \frac{ 8Q_p^{2}}{g \pi^{2} D^{4}}
  \label{eqn:demand_curve}
\end{equation}
where $H_d$ is the required pump head (in m); $H_s$ is the static head (in m); $f_D$ is the Darcy friction factor; $L_p$ is the total pipe length (in m); $D$ is the inner diameter of the pipe (in m); $\sum K$ represents the minor loss in the pipe.
The static head refers to the vertical distance between the water levels on the discharge and intake sides, which can be expressed as follows:
\begin{equation}
  H_s = \begin{cases} 
    h-h_o, &\text{ if inflow pump (water is pumped into the branch)},  \\ 
    h_o-h, &\text{ if outflow pump (water is pumped out of the branch).}
  \end{cases}
\end{equation}

Under a given operating condition, the operating point for each pump is determined by the intersection of the H-Q curve and the system demand curve. Figure \ref{fig:pump_chara}\subref{fig:pump_op_point} illustrates how the operating discharge and total head are determined from the intersection.
If the system demand curve is known, i.e., the static head is known, the pump input constraint can be obtained by finding the intersection between the system demand curve and the feasible operating region of the pump.
If the system demand curve intersects with the gray zone in Figure~\ref{fig:pump_chara}\subref{fig:pump_bound}, the pump input $N_p$ can take values within one of two disjoint ranges: $N_p=0$ or $N_p \in [{N}_{p,lb},{N}_{p,ub}]$. Otherwise, if no intersection exists, the pump should remain off ($N_p = 0$). In this case, both the lower and upper bounds of the pump input are set to zero (${N}_{p,lb}={N}_{p,ub}=0$) to maintain notational consistency. The bounds ${N}_{p,lb}$ and ${N}_{p,ub}$ may vary over time, as feasible shaft speed range depends on the static head. 
We consider that all pumps within the system are variable-speed pumps with identical parameters.

\subsection{Problem formulation\label{prob_formualation}}
With time discretization, the dynamic behaviors of the entire connected open water system can be described by the following discrete-time nonlinear form:
\begin{subequations}
  \label{eqn:watersys}
  \begin{align}
    x_{k+1} &= f(x_k, u_k, d_k) \\
    \label{eqn:watersys_y} y_k&=x_k 
  \end{align}
\end{subequations}
where $x_k \in \mathbb{X} \subset \mathbb{R}^{14}$, $u_k \in \mathbb{U} \subset \mathbb{R}^{28}$, and $d_k \in \mathbb{R}^{18}$ are the system states, control inputs, and known disturbances of the system at time instant $k$, respectively. $\mathbb X$ and $\mathbb U$ are compact sets that define the bounds for the system states and control inputs, respectively. The system states comprise the water levels of the 14 branches, denoted as $h^i$, for $i=1,\ldots, 14$. The control inputs (i.e., $u_{k}$) consist of the height of 13 weirs, denoted as $h_{w}^{i}$, for $i=1, \ldots, 13$, the shaft speeds of 11 pumps, denoted as $N_{p}^{i}$, for $i=1,\ldots, 11$, and the opening ratio of four sluice gates, denoted as $\rho_{g}^{i}$, for $i=1,\ldots, 4$. The known disturbances $d_{k}$ include the water level of each of the four connected external rivers, denoted as $h_{o}^{i}$ (in m) for, $i=1,\ldots, 4$, and the external inflow to the $i$th branch, denoted by $Q_{d}^{i}$ (in m$^{3}$/s), for $i=1,\ldots, 14$. In~\eqref{eqn:watersys_y}, $y_k$ is the vector of measured outputs. As shown in~\eqref{eqn:watersys_y}, all the 14 states are measured online during system operations.

This work aims to achieve the following control objectives: maintaining all water levels within the desired water-level zone under external disturbances, and minimizing overall energy consumption, while ensuring all state and input constraints are strictly satisfied.

\section{Mixed-integer economic zone DeePC}
\subsection{Data-enabled predictive control (DeePC)}
In this section, we review the data-enabled predictive control (DeePC) method introduced in~\cite{coulson2019Dataenabled}; this data-based DeePC framework is leveraged as the foundation of the proposed economic zone control approach for the connected open water system.

Consider a discrete-time linear time-invariant (LTI) system:
\begin{equation}
  \label{eqn:LTI_sys}
  \begin{aligned}
    x_{k+1} &= A x_k + B u_k \\
    y_k &= C x_k + D u_k
  \end{aligned}
\end{equation}
where $A \in \mathbb{R}^{n \times n}, B \in \mathbb{R}^{n \times m}, C \in \mathbb{R}^{p \times n}, D \in \mathbb{R}^{p \times m}$ are the system matrices; $x_k \in \mathbb{R}^{n}$ is the system state vector; $u_k \in \mathbb{R}^{m}$ is the control input vector; $y_k \in \mathbb{R}^{p}$ is the measured output vector.

Let $L,T \in \mathbb{Z}_{\ge 0}$ and $T\ge L$, where $\mathbb{Z}_{\ge 0}$ represents the set of non-negative integers. Define the input and output trajectories of length $T$, collected offline (denoted by the superscript $d$), as $\mathbf{u}_{[1:T]}^{d} := [ u_{1}^{d\top}, \cdots, u_{T}^{d\top} ]^{\top} \in \mathbb{R}^{mT}$ and $\mathbf{y}_{[1:T]}^{d} := [ y_{1}^{d\top}, \cdots, y_{T}^{d\top} ]^{\top} \in \mathbb{R}^{pT}$, respectively. For the input trajectory $\mathbf{u}_{[1:T]}^{d}$, the Hankel matrix of depth $L$ can be expressed as follows:
\begin{equation}
  \mathscr{H}_L(\mathbf{u}_{[1:T]}^d):=
  \begin{bmatrix}  
    u_1^d & u_2^d & \cdots & u_{T-L+1}^d \\
    u_2^d & u_3^d & \cdots & u_{T-L+2}^d \\
    \vdots & \vdots & \ddots & \vdots \\
    u_L^d & u_{L+1}^d & \cdots & u_T^d
  \end{bmatrix} 
\end{equation}
where $\mathscr{H}_L(\mathbf{u}_{[1:T]}^d) \in  \mathbb{R}^{mL \times (T-L+1)}$. Similarly, the Hankel matrix for the output trajectory can be constructed as $\mathscr{H}_L(\mathbf{y}_{[1:T]}^d)\in \mathbb{R}^{pL \times (T-L+1)}$. 
Further, consider $T_{ini}, N_c \in \mathbb{Z}_{\ge 0}$ and $L = T_{ini} + N_c$. The constructed Hankel matrices can then be partitioned into two components:
\begin{equation}
  \begin{bmatrix} U_P \\ U_F \\\end{bmatrix} := \mathscr{H}_L(\mathbf{u}_{[1:T]}^d), \quad
  \begin{bmatrix} Y_P \\ Y_F \\\end{bmatrix} := \mathscr{H}_L(\mathbf{y}_{[1:T]}^d),
\end{equation}
where $U_P \in \mathbb{R}^{mT_{ini}\times(T-L+1)}$, $U_F \in \mathbb{R}^{mN_{c}\times(T-L+1)}$, $Y_P \in \mathbb{R}^{pT_{ini}\times(T-L+1)}$ and $Y_F \in \mathbb{R}^{pN_{c}\times(T-L+1)}$. The submatrices with subscript $P$ correspond to past trajectories and are used to estimate the initial state; while the submatrices with subscript $F$ correspond to future trajectories and are used for prediction.

Data-enabled predictive control (DeePC)~\citep{coulson2019Dataenabled} utilizes Hankel matrices as an implicit representation of system dynamics to predict future trajectories and compute optimal control actions directly from offline input–output data~\citep{willems2005Note}. 
At each time instant $k$, let $\mathbf{u}_{ini,k}:=\{u\}^{k-1}_{k-T_{ini}}=[ u_{k-T_{ini}}^{\top}, \cdots, u_{k-1}^{\top} ]^{\top}$ and $\mathbf{y}_{ini,k}:=\{y\}^{k-1}_{k-T_{ini}}=[ y_{k-T_{ini}}^{\top}, \cdots, y_{k-1}^{\top} ]^{\top}$ denote the stacked past input and output trajectories, respectively; let $\hat{\mathbf{u}}_{k}:=\{\hat{u}\}^{k+N_{c}-1|k}_{k|k}$ and $\hat{\mathbf{y}}_{k}:=\{\hat{y}\}^{k+N_{c}-1|k}_{k|k}$ denote the $N_c$-step predicted input and output trajectories, respectively. DeePC solves the following constrained optimization problem~\citep{coulson2019Dataenabled}:
\begin{subequations}
  \label{eqn:opt_prob_setpoint}
  \begin{align}
    &\min_{\mathcal{G}_k, \hat{\mathbf{u}}_k, \hat{\mathbf{y}}_k} \left\|\hat{\mathbf{y}}_k-\mathbf{y}_k^r\right\|_Q^2
    +\left\|\hat{\mathbf{u}}_k\right\|_R^2 \label{eqn:sep_obj}\\
    &\text{s.t.} \;\quad
    \begin{bmatrix} U_P \\ Y_P \\ U_F \\ Y_F \\ \end{bmatrix}\mathcal{G}_k=\begin{bmatrix} \mathbf{u}_{ini,k} \\ \mathbf{y}_{ini,k} \\ \hat{\mathbf{u}}_{k} \\ \hat{\mathbf{y}}_{k} \\  \end{bmatrix}\\
    &\qquad\quad\hat{y}_{j \mid k} \in \mathbb{Y},\; \hat{u}_{j \mid k} \in \mathbb{U}, \quad j=k, \ldots, k+N_c-1
  \end{align}
\end{subequations}
where $Q \in \mathbb{R}^{pN_{c}\times pN_{c}}$ and $R\in \mathbb{R}^{mN_{c}\times mN_{c}}$ are the weighting matrices; $\mathbf{y}_k^r:=\{{y}^{r}\}^{k+N_{c}-1}_{k}\in \mathbb{R}^{pN_{c}}$ is the reference output trajectory;
$\mathbb{U} \subset \mathbb{R}^{m}$ and $\mathbb{Y} \subset \mathbb{R}^{p}$ are the input and output constraint set, respectively. 

In the online control implementation, the optimization problem \eqref{eqn:opt_prob_setpoint} is solved in a receding horizon manner. At each time instant $k$, after obtaining the optimal control input sequence $\hat{\mathbf{u}}_{k}^{*}=[\hat{u}_{k|k}^{*\top}, \hat{u}_{k+1|k}^{*\top}, \ldots ,\hat{u}_{k+N_{c}-1|k}^{*\top}]^{\top}$, only the first control input $\hat{u}_{k|k}^{*}$ is applied to the system. At next time instant $k+1$, the past trajectories $\mathbf{u}_{ini,k+1}$ and $\mathbf{y}_{ini,k+1}$ are updated with input $u_k$ and output measurement $y_k$, respectively.

\subsection{Economic zone DeePC based on lexicographic optimization\label{sec:ez-deepc}}
The two main control objectives for the connected open water system presented in Section~\ref{prob_formualation}, including maintaining water levels within the desired water-level zone (also referred to as the desired zone for brevity) and minimizing overall energy consumption, are associated with different levels of priority~\citep{horvath2022Potential}. Specifically, the controller should prioritize water-level zone tracking before addressing energy consumption minimization. Accordingly, we employ the lexicographic optimization framework~\citep{rentmeesters1996Theory}, which was also adopted in~\cite{horvath2022Potential}, to formulate two optimization problems for the two control objectives. At each sampling instant, the two optimization problems will be solved sequentially. In this way, the lower-priority control objective (i.e., minimizing the energy consumption) is addressed without compromising the results for the higher-priority control objective (i.e., maintaining water levels of the branches within the desired zone). 

Conventional DeePC approaches, e.g.,~\cite{berberich2021Datadriven} and ~\cite{coulson2019Dataenabled}, do not explicitly address zone tracking. To achieve the higher-priority control objective, we incorporate the zone tracking error, as considered in~\cite{liu2019Modelpredictive}, into the objective function of a DeePC-based controller. Additionally, considering the disjoint nature of the pump input constraints, it is natural to formulate the optimization problem as a mixed-integer programming (MIP) problem~\citep{floudas1995Nonlinear}. 

Consequently, for the higher-priority control objective of zone tracking, the first optimization problem is formulated as follows:
\begin{subequations}
  \label{eqn:opt_prob_zone}
  \begin{align}
    &\min_{\mathcal{G}_k, \hat{\mathbf{u}}_k, \hat{\mathbf{y}}_k, {\mathbf{y}}_k^{z}, \bm{\delta}_{k}} 
    \left\|\hat{\mathbf{y}}_k-\mathbf{y}_k^z\right\|_Q^2 \label{eqn:zone_obj}\\
    &\text{s.t.} \;\quad
    \begin{bmatrix} U_P \\ Y_P \\ U_F \\ Y_F \\ \end{bmatrix}\mathcal{G}_k=\begin{bmatrix} \mathbf{u}_{ini,k} \\ \mathbf{y}_{ini,k} \\ \hat{\mathbf{u}}_{k} \\ \hat{\mathbf{y}}_{k} \\  \end{bmatrix} \label{eqn:const_traj}\\
    &\qquad\quad\hat{y}_{j \mid k} \in \mathbb{Y} \label{eqn:output_bound}\\
    &\qquad\quad y^{z}_{j \mid k} \in \mathbb{Z}_{t}\label{eqn:zone_ref_con}\\
    &\qquad\quad \underline{u}^{c}_{k} \le  \hat{u}^{c}_{j \mid k} \le \overline{u}^{c}_{k} \label{eqn:inc_con}\\
    &\qquad\quad \underline{u}^{d}_{k} \odot \delta_{j \mid k} \le  \hat{u}^{d}_{j \mid k} \le \overline{u}^{d}_{k}\odot \delta_{j \mid k} \label{eqn:ind_con} \\
    &\qquad\quad \delta_{j \mid k} \in \{0,1\}^{m_p}, \quad j=k, \ldots, k+N_c-1 \label{eqn:const_bin}
  \end{align}
\end{subequations}
where $\bm{\delta}_{k}:=\{\delta\}^{k+N_c-1 \mid k}_{k \mid k} \in \mathbb{R}^{m_p N_{c}}$ is the predicted binary vector sequence; 
$\mathbf{y}_k^z:=\{{y}^{z}\}^{k+N_{c}-1|k}_{k|k} \in \mathbb{R}^{pN_{c}}$ is the reference output trajectory; $\mathbb{Y} \subset \mathbb{R}^{p}$ is the output constraint set; $\mathbb{Z}_{t} \subset \mathbb{R}^{p}$ is the control target zone set; $\odot$ is Hadamard (element-wise) product. The input is partitioned into two components as $\hat{u}_{j \mid k} = [\hat{u}^{c \top}_{j \mid k}, \hat{u}^{d \top}_{j \mid k}]^{\top}$. The vector $\hat{u}^{d}\in \mathbb{R}^{m_p}$ represents the pump inputs $[{N}_{p}^{1}, \ldots ,{N}_{p}^{m_p}]^{\top}$ in the prediction horizon, where $m_p$ is the dimension of pump inputs. Due to the on/off operating modes of the pumps, these inputs are subject to disjoint constraints. The vector $\hat{u}^{c}\in \mathbb{R}^{m_w+m_g}$ represents the continuous weir and gate inputs $[{h}_{w}^{1}, \ldots ,{h}_{w}^{m_w}, {\rho}_{g}^{1}, \ldots ,{\rho}_{g}^{m_g}]^{\top}$ in the prediction horizon, where $m_w$ and $m_g$ are the dimensions of weir inputs and gate inputs, respectively.

In the optimization problem, \eqref{eqn:zone_obj} aims to minimize the zone tracking loss. \eqref{eqn:output_bound} defines the constraint on system output $y$, and \eqref{eqn:zone_ref_con} restricts the reference output $y^{z}$ to a control target zone for zone tracking purposes. Note that the output constraint in \eqref{eqn:output_bound} reflects system safety requirements, while the control target zone in~\eqref{eqn:zone_ref_con} is designed to balance zone-tracking performance and energy efficiency under complex system dynamics and external disturbances. The detailed procedure for determining the control target zone is provided in Section~\ref{sec:BO}. In \eqref{eqn:inc_con}, the continuous input $\hat{u}^{c}_{j \mid k}$ is constrained by upper bound $\overline{u}^{c}_{k}$ and lower bound $\underline{u}^{c}_{k}$. In \eqref{eqn:ind_con}, each element of the input vector $\hat{u}^{d}_{j \mid k}$ is constrained individually: for the $i$th component, the bounds $\overline{u}^{d,i}_{k}$ and $\underline{u}^{d,i}_{k}$ are applied only when $\delta^{i} = 1$, otherwise $\hat{u}^{d,i}_{j \mid k}$ must be 0, for $i=1, \ldots ,m_{p}$. In \eqref{eqn:const_bin}, the auxiliary binary vector $\delta \in {\{0,1\}}^{m_p}$ represents the on/off status of the pumps. 
The bounds on the inputs in \eqref{eqn:inc_con} and \eqref{eqn:ind_con} are determined based on system output $y_{k}$ and disturbance $d_{k}$, as discussed in Section \ref{chap:pump_model}. These input constraints define a time-varying input constraint set $\mathbb{U}_k \subset \mathbb{U} \in \mathbb{R}^{m}$, where $\mathbb{U}$ represents the set of all physically feasible inputs. Although the input bounds may vary with the predicted states, $\mathbb{U}_{k}$ is considered to remain constant throughout each control horizon. 

The objective function of set-point tracking DeePC \citep{coulson2019Dataenabled} in \eqref{eqn:sep_obj} includes a penalty term on the control input. However, the zone DeePC objective function in \eqref{eqn:zone_obj} does not include such a penalty term, since the control inputs will naturally vary with fluctuating river water levels and disturbance inflows. Under these conditions, specifying a fixed input reference for the controller to track would be inappropriate.

We note that the optimal control input obtained by solving~\eqref{eqn:opt_prob_zone} is not directly applied to the connected open water system~\eqref{eqn:watersys}. Instead, the optimal zone tracking performance obtained from solving~\eqref{eqn:opt_prob_zone} is incorporated to constrain the pursuit the lower-priority control objective, that is, to minimize the energy consumption without compromising the zone tracking performance already achieved through~\eqref{eqn:opt_prob_zone}. Specifically, the optimized predicted output $\hat{\mathbf{y}}^{*}_{k}$ and zone reference output ${\mathbf{y}}^{z*}_{k}$ obtained from solving~\eqref{eqn:opt_prob_zone} are used to compute the minimum zone tracking cost $zc^{*}= \left\|\hat{\mathbf{y}}_k^{*}-\mathbf{y}_k^{z*}\right\|_Q^2$, which constrains the zone tracking cost when minimizing the energy consumption. 

Consequently, the second optimization problem, which corresponds to the lower-level priority control objective of energy consumption minimization, is formulated as:
\begin{subequations}
  \label{eqn:opt_prob_eco}
  \begin{align}
    &\min_{\mathcal{G}_k, \hat{\mathbf{u}}_k, \hat{\mathbf{y}}_k, {\mathbf{y}}_k^{z}, \bm{\delta}_{k}} 
    \sum_{j=k}^{k+N_{c}-1} l_{e}(\hat{y}_{j|k}, \hat{u}_{j|k}, {d}_{j|k}) \label{eqn:eco_obj} \\
    &\text{s.t.} \;\;\quad \left\|\hat{\mathbf{y}}_k-\mathbf{y}_k^z\right\|_Q^2 \le zc^{*} \label{eqn:lexic_con}\\
    &\qquad\quad \eqref{eqn:const_traj} - \eqref{eqn:const_bin}
  \end{align}
\end{subequations}
where ${d}_{j|k}$ is the predicted external disturbance of time instant $j$ at time instant $k$. The term $l_{e}$ in~\eqref{eqn:eco_obj} represents energy consumption, computed as the energy usage of the pumps in one sampling period as follows:
\begin{equation}
  l_{e}(\hat{y}_{j|k}, \hat{u}_{j|k}, {d}_{j|k}) = 
   \sum_{i=1}^{m_p}P_{p,j}^{i} \Delta t
\end{equation}
where $P_{p,j}^{i}$ is the power consumption of $i$th pump at time instant $j$ and $\Delta t$ is the system sampling period, which is computed as in~\eqref{eqn:pump_energy}. \eqref{eqn:lexic_con} ensures that the zone tracking cost is no greater than the optimal value obtained from~\eqref{eqn:opt_prob_zone}. Since the future disturbances are unknown at current time instant, we assume that the predicted disturbance ${d}_{j|k}$ remains constant in the prediction horizon. 

At time instant $k \in \mathbb{Z}_{\ge 0}$, the two optimization problems in~\eqref{eqn:opt_prob_zone} and~\eqref{eqn:opt_prob_eco} are solved sequentially. The optimal input $u^{*}_{k\mid k}$ obtained from solving \eqref{eqn:opt_prob_eco} is applied to the connected open water system in~\eqref{eqn:watersys}.

Compared with conventional DeePC~\citep{berberich2021Datadriven, coulson2019Dataenabled}, the proposed approach incorporates both water-level zone tracking and energy optimization objectives. Accordingly, a mixed-integer programming (MIP) problem is formulated to handle the disjoint operating regions of the pumps, and lexicographic optimization~\citep{rentmeesters1996Theory} is employed to systematically manage the multiple control objectives. Additionally, a Bayesian optimization (BO) module, as described in Section~\ref{sec:BO}, is integrated to automatically determine the optimal control target zone.

The proposed framework can be extended to more complex and comprehensive representations of open water systems. Particularly, by spatially discretizing the S--V equations (e.g., using a staggered-grid scheme as in~\cite{stelling2003Staggered}), a surrogate ordinary differential-equation (ODE) formulation can be obtained. This allows the proposed control framework to be applied to higher-fidelity hydrodynamic models such as the S--V equations.

\subsection{Regularization and dimension reduction}

Although DeePC was originally developed for linear time-invariant (LTI) systems, recent studies in \cite{shang2024Willems} and \cite{xiong2025Dataenabled} have shown that it can be extended to nonlinear systems. The underlying idea is that a nonlinear system can be accurately approximated using a linear form in a lifted (high-dimensional) feature space, while preserving the same input-output behavior as the underlying nonlinear system. 
To handle nonlinearities and disturbances present in the connected open water system, it is beneficial to employ appropriate regularization techniques~\citep{dorfler2023Bridging} and collect sufficiently long offline trajectories~\citep{shang2024Willems}. However, extending the trajectory length increases the complexity of the resulting optimization problem, which leads to a higher computational burden~\citep{zhang2023Dimension}. To mitigate this issue, we adopt the $\gamma$-DDPC algorithm~\citep{breschi2023Datadriven, breschi2023Uncertaintyaware}, which integrates both dimension reduction and regularization.

Let us denote 
\begin{equation}
  \mathbf{z}_{ini,k}= \begin{bmatrix} \mathbf{u}_{ini,k} \\ \mathbf{y}_{ini,k} \\\end{bmatrix},\quad
  Z_{P} = \begin{bmatrix} U_{P} \\ Y_{P} \\\end{bmatrix}.
\end{equation}
Let the offline trajectory be sufficiently long such that the input-output Hankel matrix $[Z_{P}^{\top}, U_{F}^{\top}, Y_{F}^{\top}]^{\top}$ has more columns than rows. With LQ decomposition \citep{golub2013Matrix}, the Hankel matrix can be expressed as the product of a lower-triangular matrix and a row-orthogonal matrix, and \eqref{eqn:const_traj} can be rewritten as follows:
\begin{equation}
  \label{eqn:deepc_traj_simp}
  \begin{bmatrix} \mathbf{z}_{ini,k} \\ \hat{\mathbf{u}}_{k} \\ \hat{\mathbf{y}}_{k} \\\end{bmatrix} = 
  \begin{bmatrix} Z_{P} \\ U_F \\ Y_F \\\end{bmatrix}\mathcal{G}_k = 
  \begin{bmatrix} L_{11} & \bm{0} & \bm{0} \\ L_{21} & L_{22} & \bm{0} \\ L_{31} & L_{32} & L_{33} \\\end{bmatrix}
  \begin{bmatrix} Q_1 \\ Q_2 \\ Q_3 \\\end{bmatrix} \mathcal{G}_k \triangleq \begin{bmatrix} L_{11} & \bm{0} & \bm{0} \\ L_{21} & L_{22} & \bm{0} \\ L_{31} & L_{32} & L_{33} \\\end{bmatrix}
  \begin{bmatrix} \gamma_{1,k} \\ \gamma_{2,k} \\ \gamma_{3,k} \\\end{bmatrix},
  \end{equation}
where $\gamma_{i,k}=Q_i \mathcal{G}_k$, for $i=1,2,3$. Since $\mathbf{z}_{ini,k}$ is known at each time instant $k$, $\gamma_{1,k}^{*}$ can be computed with $\gamma_{1,k}^{*} = L_{11}^{\dag} \mathbf{z}_{ini,k}$. Therefore, instead of optimizing over $\mathcal{G}_k$, we optimize over
\begin{equation}
  \gamma_k = 
   \begin{bmatrix} \gamma_{2,k} \\ \gamma_{3,k} \\\end{bmatrix}
   \in \mathbb{R}^{(m+p)N_{c}},
\end{equation}
which is independent of both the offline trajectory length $T$ and the online past trajectory length $T_{ini}$.
By introducing regularization terms on $\gamma_{k}$, we reformulate the optimization problem for zone tracking \eqref{eqn:opt_prob_zone} as follows:
\begin{subequations}
  \label{eqn:opt_prob_reg_zone}
  \begin{align}
    &\min_{\gamma_k, \hat{\mathbf{u}}_k, \hat{\mathbf{y}}_k, \bm{\delta}_{k}} 
    \left\|\hat{\mathbf{y}}_k-\mathbf{y}_k^z\right\|_Q^2+ \beta_{2,z} \left\| \gamma_{2,k} \right\|^2 + \beta_{3,z} \left\| \gamma_{3,k} \right\|^2\\
    &\text{s.t.} \;\quad 
    \begin{bmatrix} \hat{\mathbf{u}}_{k} \\ \hat{\mathbf{y}}_{k} \\  \end{bmatrix} = 
    \begin{bmatrix} L_{22} & \bm{0} \\ L_{32} & L_{33} \\\end{bmatrix}
    \begin{bmatrix} \gamma_{2,k} \\ \gamma_{3,k} \\\end{bmatrix} +
    \begin{bmatrix} L_{21} \\ L_{31} \\\end{bmatrix} \gamma_{1,k}^{*}\\
    &\qquad\quad \eqref{eqn:output_bound} - \eqref{eqn:const_bin} 
  \end{align}
\end{subequations}
where $\beta_{2,z}$ and $\beta_{3,z}$ are the weighting coefficients.

Similarly, the optimization problem for energy consumption minimization is formulated as follows:
\begin{subequations}
  \label{eqn:opt_prob_reg_eco}
  \begin{align}
    &\min_{\gamma_k, \hat{\mathbf{u}}_k, \hat{\mathbf{y}}_k, \bm{\delta}_{k}} 
    l_{e}(\hat{\mathbf{y}}_{k}, \hat{\mathbf{u}}_{k}, {d}_{k})+ \beta_{2,e} \left\| \gamma_{2,k} \right\|^2 + \beta_{3,e} \left\| \gamma_{3,k} \right\|^2\\
    &\text{s.t.} \;\quad 
    \begin{bmatrix} \hat{\mathbf{u}}_{k} \\ \hat{\mathbf{y}}_{k} \\  \end{bmatrix} = 
    \begin{bmatrix} L_{22} & \bm{0} \\ L_{32} & L_{33} \\\end{bmatrix}
    \begin{bmatrix} \gamma_{2,k} \\ \gamma_{3,k} \\\end{bmatrix} +
    \begin{bmatrix} L_{21} \\ L_{31} \\\end{bmatrix} \gamma_{1,k}^{*}\\
    &\qquad\quad \left\|\hat{\mathbf{y}}_k-\mathbf{y}_k^z\right\|_Q^2 \le zc^{*}\\
    &\qquad\quad \eqref{eqn:output_bound} - \eqref{eqn:const_bin} 
  \end{align}
\end{subequations}
where $\beta_{2,e}$ and $\beta_{3,e}$ are the weighting coefficients.

\section{Determination of control target zone\label{sec:BO}}

To achieve satisfactory water-level zone tracking performance through solving~\eqref{eqn:opt_prob_reg_zone}, the control target zone $\mathbb{Z}_{t}$ in~\eqref{eqn:zone_ref_con} needs to be appropriately determined. Due to inherent system nonlinearities and the presence of known disturbances, the input-output Hankel matrix in \eqref{eqn:const_traj} may fail to accurately capture the actual system dynamics, which leads to mismatches between the predicted and actual output trajectories. If the desired water-level zone $\mathbb{Z}_{d}$ is directly used as the control target zone set $\mathbb{Z}_{t}$ in~\eqref{eqn:zone_ref_con}, the water levels may violate the desired zone. On the other hand, imposing an overly tight control target zone may lead to unnecessary increases in energy consumption. To balance this trade-off, Bayesian optimization (BO)~\citep{frazier2018Tutorial} is employed to determine an appropriate control target zone set $\mathbb{Z}_{t}$.

BO is an approach for solving black-box optimization problems with time-consuming or expensive evaluations~\citep{frazier2018Tutorial}. It constructs a probabilistic surrogate model of the objective function using Gaussian process regression (GPR)~\citep{williams2006gaussian}, which estimates both the function value and its uncertainty. In each evaluation step, an acquisition function is employed to guide the selection of the next evaluation point based on the surrogate model. After evaluating the objective function at the selected point, the result is added to the dataset, and the surrogate model is updated~\citep{frazier2018Tutorial}. BO has been widely applied for hyperparameter tuning in deep learning \citep{snoek2015Scalable,wang2023Recent}, and for tuning model predictive controllers \citep{lu2021Bayesian,sorourifar2021Datadriven}.

We consider the case where the desired zone is time-invariant and has uniform width across all branches. Let $y_{c} \in \mathbb{R}^{p}$ be the center of the zone, $\Delta y \in \mathbb{R}_{>0}$ represent half the width of each zone. The desired zone set is defined as:
\begin{equation}
  \mathbb{Z}_{d} = \{y\in \mathbb{R}^{p}: y_{c}-\Delta y \cdot \bm{1}_{p} \le y \le y_{c} + \Delta y \cdot \bm{1}_{p}\}.
\end{equation}
The control target zone shares the same center $y_{c}$ as the desired zone. The upper and lower bounds of the control target zone $\mathbb{Z}_{t}$ are symmetrically contracted from those of the desired zone $\mathbb{Z}_{d}$ using a zone contraction rate $\alpha \in [0, 1]$. The resulting control target zone is defined as:
\begin{equation}
  \label{eqn:control_target_zone}
  {\mathbb{Z}_{t}} = \{y\in \mathbb{R}^{p}: y_{c}-\alpha \Delta y \cdot \bm{1}_{p} \le y \le y_{c} + \alpha \Delta y \cdot \bm{1}_{p}\}.
\end{equation}

To balance zone tracking performance and energy consumption, we employ BO~\citep{frazier2018Tutorial} to solve the following problem:
\begin{equation}
  \max_{\alpha \in \mathbb{A}}\varphi(\alpha)
\end{equation}
where $\mathbb{A}=\{\alpha \in \mathbb{R}: 0\le \alpha \le 1\}$ denotes the feasible set of $\alpha$, and $\varphi(\alpha)$ is the objective function. $\varphi(\alpha)$ can be evaluated by performing closed-loop control with the given $\alpha$ and computing the weighted sum of the zone tracking cost and the energy consumption in the evaluation time period:
\begin{equation}
  \label{eqn:bo_obj}
  \varphi(\alpha) = - \sum_{k=T_{ini}}^{T_{b}+T_{ini}-1} \left( \min_{y^{z}_{k} \in \mathbb{Z}_{d}}\| y_{k} - y^{z}_{k} \|_{1} + \lambda_{b} l_{e}(y_k,u_k,d_k) \right)
\end{equation}
where $T_{b}\in \mathbb{Z}_{\ge 0}$ is the evaluation horizon length; $\lambda_{b}\in \mathbb{R}_{\ge 0}$ is a tunable weighting parameter; $l_{e}(y_k,u_k,d_k) = \sum_{i=1}^{m_p}P_{p,k}^{i} \Delta t$ is the energy consumption at time instant $k$. Note that the zone tracking cost is determined based on the predefined desired zone $\mathbb{Z}_{d}$ instead of the control target zone $\mathbb{Z}_{t}$. In each evaluation step, the disturbance trajectory $\{d\}_{0}^{T_{b}-1}$ is set to be the same. The initial state $x_0$ is randomly selected from a specified range in each evaluation step to make sure that the optimized zone can perform robustly under different initial conditions.

In Bayesian optimization, the objective function~\eqref{eqn:bo_obj} is modeled as a Gaussian process (GP). Let $\alpha_{1:w}=[\alpha_{1},\cdots,\alpha_{w}]^{\top}$ be the sample points, and $\varphi(\alpha_{1:w})=[\varphi(\alpha_{1}),\cdots,\varphi(\alpha_{w})]^{\top}$ be the corresponding objective function values. The objective function values are assumed to be drawn from a multivariate normal prior probability distribution as follows:
\begin{equation}
  \varphi(\alpha_{1:w}) \sim \mathcal{N}(\mu, \Sigma)
\end{equation}
where $\mu = \mu_0(\alpha_{1:w}) \in \mathbb{R}^{w}$ is the mean vector and $\Sigma = \Sigma_0(\alpha_{1:w},\alpha_{1:w}) \in \mathbb{R}^{w\times w}$ is the covariance matrix with $[\Sigma]_{i,j}=\Sigma_0(\alpha_{i},\alpha_{j})$. The prior mean is set to zero, i.e., $\mu_0(\alpha)=0$, and $\Sigma_0(\alpha,\alpha')$ is the prior covariance function, for which we adopt the M{\`a}tern kernel~\citep{frazier2018Tutorial}.

Since the initial state is randomly selected, the objective function~\eqref{eqn:bo_obj} may have different values even at identical input parameters. Following~\cite{frazier2018Tutorial}, an additional diagonal noise term is added to the covariance matrix to account for this variability. The prior distribution can be expressed as $\varphi(\alpha_{1:w}) \sim \mathcal{N}(\mu, \Sigma + \sigma_{o}^{2} I_{w})$, where $\sigma_{o}^{2}$ is the observation noise variance. For any other sample point $\alpha$, the objective function value $\varphi(\alpha)$ and the previously observed data $\varphi(\alpha_{1:w})$ follow a joint Gaussian distribution under the GP prior distribution. The conditional distribution of $\varphi(\alpha)$ given the observed data can be computed using the Bayes' rule~\citep{williams2006gaussian,frazier2018Tutorial}:
\begin{equation}
  \label{eqn:bo_posterior}
  \begin{aligned}
    \varphi(\alpha)|\varphi(\alpha_{1:w}) &\sim \mathcal{N}(\mu_{w}(\alpha), \sigma_{w}(\alpha))\\
    \mu_{w}(\alpha) &= \mu_0(\alpha)+\Sigma_0(\alpha,\alpha_{1:w})(\Sigma+\sigma_{o}^{2} I_{w})^{-1}(\varphi(\alpha_{1:w})-\mu)\\
    \sigma_{w}(\alpha) &= \Sigma_0(\alpha, \alpha) - \Sigma_0(\alpha,\alpha_{1:w})(\Sigma+\sigma_{o}^{2} I_{w})^{-1}\Sigma_0(\alpha_{1:w}, \alpha).
  \end{aligned}
\end{equation}

At the beginning, a small number of samples are needed to initialize the optimization process. For the first $W_{ini}$ evaluations, $\alpha$ is selected from a predefined set $\mathcal{A}_{ini}$ based on a uniform sampling scheme. 
In the subsequent evaluations, an acquisition function is constructed based on the conditional distribution \eqref{eqn:bo_posterior} to guide the selection of $\alpha$. Here, we use the upper confidence bound (UCB)~\citep{brochu2010Tutorial} as the acquisition function:
\begin{equation}
  \label{eqn:acq_func}
  \mathrm{UCB}(\alpha) = \mu_{w}(\alpha) + \kappa \sigma_{w}(\alpha)
\end{equation}
where $\kappa \in \mathbb{R}_{\ge 0}$ is the weighting factor to balance exploration and exploitation. Given the previous observations $(\alpha_{1:w}, \varphi(\alpha_{1:w}))$, the next candidate point for evaluation is obtained by solving the following problem:
\begin{equation}
  \label{eqn:ucb}
  \alpha_{w+1} = \max_{\alpha \in \mathbb{A}} \mathrm{UCB}(\alpha).
\end{equation}
After solving \eqref{eqn:ucb}, the objective function at $\alpha_{w+1}$ is evaluated, and $(\alpha_{w+1}, \varphi(\alpha_{w+1}))$ is added to the dataset to update the surrogate model for the next evaluation. 

\begin{figure}[t!]
  \centering
  \includegraphics[width=0.93\textwidth]{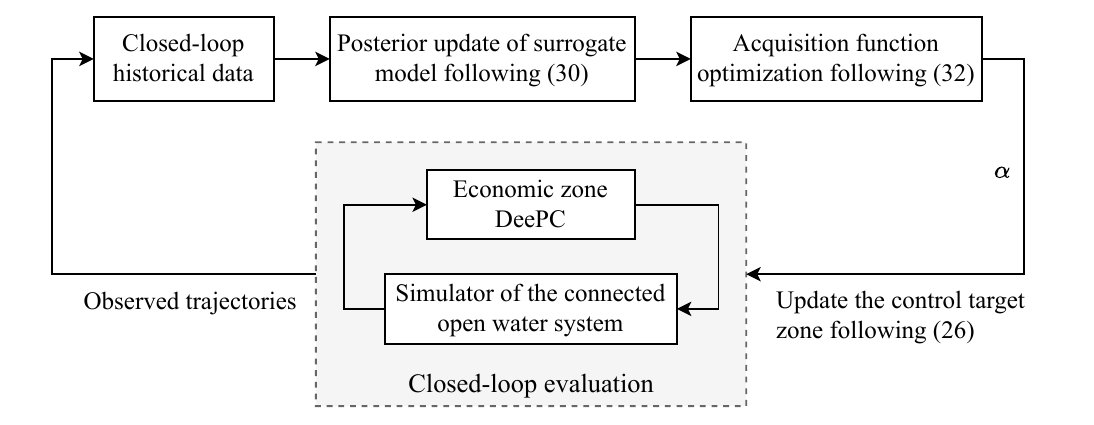}
  \caption{An illustrative diagram of Bayesian optimization applied to control target zone determination.}
  \label{fig:bo_diagram}
\end{figure}

The Bayesian optimization process continues until a predefined maximum number of evaluations, $W_{\max}$, is reached. Finally, the point with the highest posterior mean is selected:
\begin{equation}
  \label{eqn:bo_res}
  \alpha^{*}=\max_{\alpha \in \mathbb{A}}\mu_{W_{\max}}(\alpha).
\end{equation} 
This optimal value $\alpha^{*}$ is then adopted as the final zone contraction rate, which is substituted into~\eqref{eqn:control_target_zone} to determine the control target zone $\mathbb{Z}_{t}$. During online implementation, the economic zone DeePC controller \eqref{eqn:opt_prob_reg_zone} and \eqref{eqn:opt_prob_reg_eco} operates with this fixed control target zone, as determined offline through BO.
An overview of the Bayesian optimization procedure is shown in Figure~\ref{fig:bo_diagram}. The implementation of the BO-based control target zone selection process is summarized in Algorithm~\ref{alg:1}.

\begin{algorithm}[!t]
  \caption{BO-based control target zone determination}
  \label{alg:1}
  \begin{spacing}{1.3}
  \begin{algorithmic}[1]
  \Require Offline input/output data $\mathbf{u}^{d}_{T}$, $\mathbf{y}^{d}_{T}$; initial trajectory length $T_{ini}$; prediction horizon $N_c$; admissible input set $\mathbb{U}$; output constraint set $\mathbb{Y}$; output weighting matrix $Q$; weighting terms $\beta_{2,z}$, $\beta_{3,z}$, $\beta_{2,e}$, $\beta_{3,e}$, $\lambda_e$, $\lambda_{b}$; zone center $y_{c}$; half of zone width $\Delta y$; number of initial BO evaluations $W_{ini}$; maximum BO evaluation steps $W_{\max}$; BO evaluation horizon $T_{b}$; BO initial parameter set $\mathcal{A}_{ini}$
  \Ensure Optimal zone contraction rate $\alpha^{*}$
  
  \State Construct Hankel matrices and perform dimension reduction following \eqref{eqn:deepc_traj_simp}
  
  \For{$i = 1,2, \ldots, W_{max}$}
    \If{$i \leq W_{ini}$}
      \State Select candidate parameter $\alpha_i$ from the predefined set $\mathcal{A}_{ini}$
    \Else
      \State Select $\alpha_i$ using acquisition function following \eqref{eqn:ucb}
    \EndIf
    \State Initialize $\mathbf{u}_{ini,T_{ini}}$, $\mathbf{y}_{ini,T_{ini}}$
    \State Construct control target zone $\mathbb{Z}_{t}$ with $\alpha_{i}$ following \eqref{eqn:control_target_zone}
    \For{$k = T_{ini}+1,T_{ini}+2,\ldots,T_{ini}+T_{max}$}
      \State Solve \eqref{eqn:opt_prob_reg_zone} and \eqref{eqn:opt_prob_reg_eco} sequentially for the optimal input sequence $\hat{\mathbf{u}}_{k}^{*}$
      \State Apply control input $u_{k}=\hat{u}^{*}_{k|k}$ to the system \eqref{eqn:watersys}
      \State Update $\mathbf{u}_{ini,k+1}:=\{u\}^{k}_{k-T_{ini}+1}$, $\mathbf{y}_{ini,k+1}:=\{y\}^{k}_{k-T_{ini}+1}$
    \EndFor
    \State Compute the BO objective $\varphi(\alpha_i)$ following~\eqref{eqn:bo_obj}
    \State Perform posterior update of the surrogate model following~\eqref{eqn:bo_posterior}
  \EndFor
  \State Determine optimal parameter $\alpha^{*}$ following \eqref{eqn:bo_res}
   
  \end{algorithmic}
  \end{spacing}
\end{algorithm}

We note that in real-world applications, it is typically not feasible to reproduce identical disturbance trajectories across multiple experiments. This poses a challenge for Bayesian optimization, which relies on consistent evaluation conditions when comparing different parameter choices~\citep{frazier2018Tutorial}. To address this issue, we can perform the BO-based parameter tuning via simulations based on a high-fidelity simulator that accurately captures the dynamics of the underlying connected open water system. This enables repeatable and reliable evaluations during the optimization process. 
After identifying the optimal parameter $\alpha^*$ from simulation, it can be applied directly in the real system during online operation. This treatment will help avoid the impracticality of repeated field experiments.

\section{Results}

\subsection{Settings}
The parameters of the connected open water system \eqref{eqn:watersys} are listed in Table~\ref{tab:sys_param}. The backwater area $A$ and weir crest width $w_{w}$ are adopted or adapted from~\cite{horvath2019Closedloop}. The discharge coefficients $C_{dw}$ and $C_{dg}$ are adopted from~\cite{horvath2022Potential} and~\cite{bos1989Discharge}, respectively. 
For the minor loss term $\sum K$ in \eqref{eqn:demand_curve}, only the loss of kinetic energy at the pipe exit is considered, resulting in $\sum K = 1.0$~\citep{potter2011mechanics}.

{
  \renewcommand{\arraystretch}{1.2}
  \begin{table}[!t]
    \centering
    \caption{Parameters of the connected open water system.}
    \label{tab:sys_param}
    \begin{tabular}{p{5.5cm}p{10cm}}
      \toprule
      \textbf{Parameter} & \textbf{Value} \\ \midrule
      \textbf{Branch} &  \\ 
      Backwater area $A$ for each of the 14 branches (m$^{2}$) & 141682; 26416; 47601; 43848; 47712; 76457; 270461; 55691; 99111; 436163; 103840; 210146; 150000; 900000\\
      \midrule
      \textbf{Weir} &  \\ 
      Discharge coefficient $C_{dw}$ (-) & 0.61 \\
      Crest width $w_{w}$ for each of the 13 weirs (m) & 6.0; 6.0; 6.0; 6.0; 6.0; 6.0; 5.94; 5.94; 6.0; 9.5; 9.5; 12; 20\\
      \midrule
      \textbf{Pump} &  \\ 
      Darcy friction factor $f_{D}$ (-) & 0.013 \\
      Total pipe length $L_{p}$ (m) & 50\\
      Pipe inner diameter $D$ (m) & 1.8288 \\
      Minor losses $\sum K$ (-) & 1.0 \\
      Power consumption curve coefficients $a_1,a_2,a_3,a_4$ (-) & -1.81; 19.72; -83.06; 506.15 \\  \midrule
      \textbf{Sluice gate} &  \\ 
      Discharge coefficient $C_{dg}$ (-) & 0.61 \\
      Width $w_g$ for each of the four gates (m) & 5.0; 6.0; 6.5; 18\\
      Maximum gate opening $L_{g}$ for each of the four gates (m) & 0.6; 0.6; 0.6; 0.6\\
      \bottomrule
    \end{tabular}
  \end{table}
}

In simulation, the sampling period is set to $\Delta t = 0.5$ hour. 
The centers of the output constraint ranges and the desired zone are the same, and are shown in Table~\ref{tab:output_constraints}. The output constraints require the water level of each branch to remain within $\pm 0.3$ m of the center value, while the desired zone for each branch, being set to $\pm 0.1$ m, is narrower. Note that output constraints ensure basic system safety, while the desired zone aims for appropriate system operation.

The physical constraints on the control inputs imposed by the limited capacity of the hydraulic structures define the admissible input set $\mathbb{U}$, as summarized in Table~\ref{tab:input_constraint}. Note that the maximum and minimum pump input values apply only when the pump is operational. When the pump is shut down, the shaft speed is set to $N_{p}=0$.

{
  \renewcommand{\arraystretch}{1.2}
  \begin{table}[!t]
    \centering
    \caption{Center of the output constraint range and the desired zone for each of the 14 branches.}
    \label{tab:output_constraints}
    \begin{tabular}{cccccccccccccc}
    \toprule
    $h^1$ & $h^2$ & $h^3$ & $h^4$ & $h^5$ & $h^6$ & $h^7$ & $h^8$ & $h^9$ & $h^{10}$ & $h^{11}$ & $h^{12}$ & $h^{13}$ & $h^{14}$ \\ \midrule
    9.0  & 8.6  & 8.16  & 8.0 & 7.3  & 6.68  & 5.85  & 5.6  & 4.6  & 3.85  & 3.0  & 2.1  & 1.45  & 0.8 \\ \bottomrule
    \end{tabular}
  \end{table}
}

{
  \renewcommand{\arraystretch}{1.2}
  \begin{table}[!t]
    \centering
    \caption{Constraints on the control inputs $u$.}
    \label{tab:input_constraint}
    \begin{tabular}{cccccccccccccccc}
      \toprule
      & $h_{w}^{1}$ & $h_{w}^{2}$ & $h_{w}^{3}$ & $h_{w}^{4}$ & $h_{w}^{5}$ & $h_{w}^{6}$ & $h_{w}^{7}$ & $h_{w}^{8}$ & $h_{w}^{9}$ & $h_{w}^{10}$ & $h_{w}^{11}$ & $h_{w}^{12}$ & $h_{w}^{13}$ & $\rho_{g}$ & $N_{p}$ \\ \midrule
      $\max$ & 11.5 & 11.0 & 10.0 & 9.5 & 9.0 & 8.0 & 7.5 & 6.5 & 5.5 & 4.5 & 4.0 & 3.5 & 2.5 & 1.0 & 250\\ 
      $\min$ & 7.8 & 7.5 & 7.0 & 6.5 & 6.0 & 5.0 & 3.5 & 3.0 & 2.5 & 1.5 & 0.8 & 0.6 & 0.4 & 0.0 & 120\\ 
      \bottomrule
    \end{tabular}
  \end{table}
}

\begin{figure}[t!]
  \centering
  \includegraphics[width=1.0\textwidth]{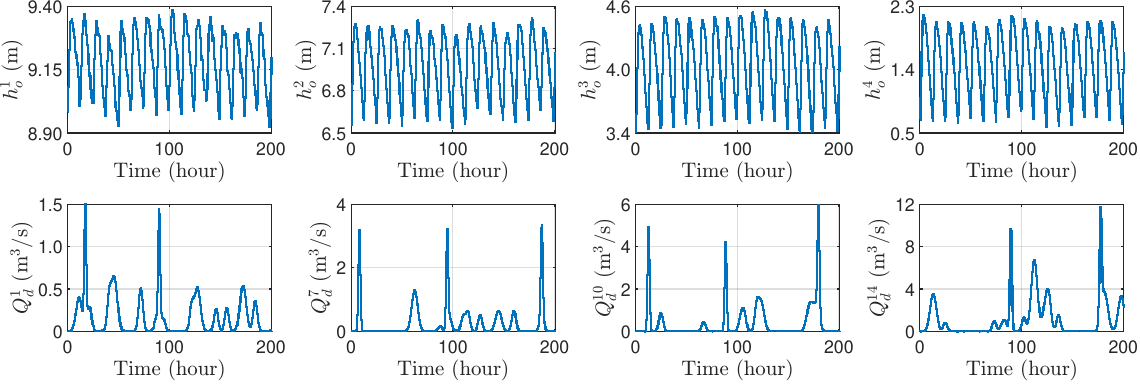}
  \caption{A representative set of disturbance trajectories including the water levels of external rivers and the disturbance inflow to branches 1, 7, 10, and 14.}
  \label{fig:dist_traj}
\end{figure}

The system disturbances comprise the water levels of the connected external rivers and the disturbance inflows to the 14 branches. These disturbances are measured at each sampling instant. 
The water levels of the external rivers are primarily influenced by tides, with stronger tidal effects observed in those connected to the downstream branches. The disturbance inflows to the 14 branches are primarily driven by weather-related factors~\citep{lund2018Model}, following the data patterns reported in~\cite{horvath2022Potential}. These inflows are assumed to be independent across the 14 branches. Furthermore, the peak inflow for each branch is assumed to be proportional to its backwater area. A representative set of disturbance trajectories is illustrated in Figure \ref{fig:dist_traj}.

Open-loop step input signals are used to excite the system during data collection. The step levels are randomly selected and updated every 10 sampling periods. To ensure feasibility, the inputs are clipped at each time step to satisfy the state-dependent input constraints.
Since the system exhibits nonlinear dynamics, data collected far from the desired zone may poorly represent the local behavior around the desired zone, which compromises the effectiveness of the controller.
To address this issue, corrective interventions are made during open-loop data collection to maintain the system output within $\pm 0.5$ m of the center of the desired zone. Specifically, when the water level of a branch becomes too high and its adjacent branch has a lower water level, the weir height between the two branches is gradually lowered to increase the flow toward the downstream side. For branches equipped with pumps and gates, operational directions are further constrained, that is, only pumps and gates that discharge water outward are allowed to operate. Conversely, when the water level of a branch is too low, the weir height is gradually increased, and only inward pump and gate flows are permitted.

\subsection{Bayesian optimization-based control target zone determination}
For the economic zone DeePC controller~\eqref{eqn:opt_prob_reg_zone} and~\eqref{eqn:opt_prob_reg_eco}, the parameters are chosen as: $T=12000$; $T_{ini}=15$; $N_{c}=5$; $Q=5\times I_{70}$; $\beta_{2,z}=0.5$; $\beta_{3,z}=100$; $\beta_{2,e}=5 \times 10^{3}$; $\beta_{3,z}=10^{6}$. The Knitro solver~\citep{byrd2006Knitro} is employed to solve the formulated mixed-integer nonlinear programming problems.

In the Bayesian optimization (BO) process, the evaluation length $T_{b}$ is 200 sampling periods, and the maximum evaluation steps is $W_{\max}=16$. $\lambda_{b}$ in the objective function~\eqref{eqn:bo_obj} is set to $2.5\times 10^{-4}$. The observation noise variance is set to $\sigma_{o}^{2}=0.35^{2}$. The number of initial BO evaluations is set to $W_{ini}=3$, and the initial parameter set is set to $\mathcal{A}_{ini}=\{1.0, 0.5, 0.0\}$. The weighting factor in the acquisition function \eqref{eqn:acq_func} is set to $\kappa=2.576$. 

\begin{figure}[t!]
  \centering
  \includegraphics[width=1.0\textwidth]{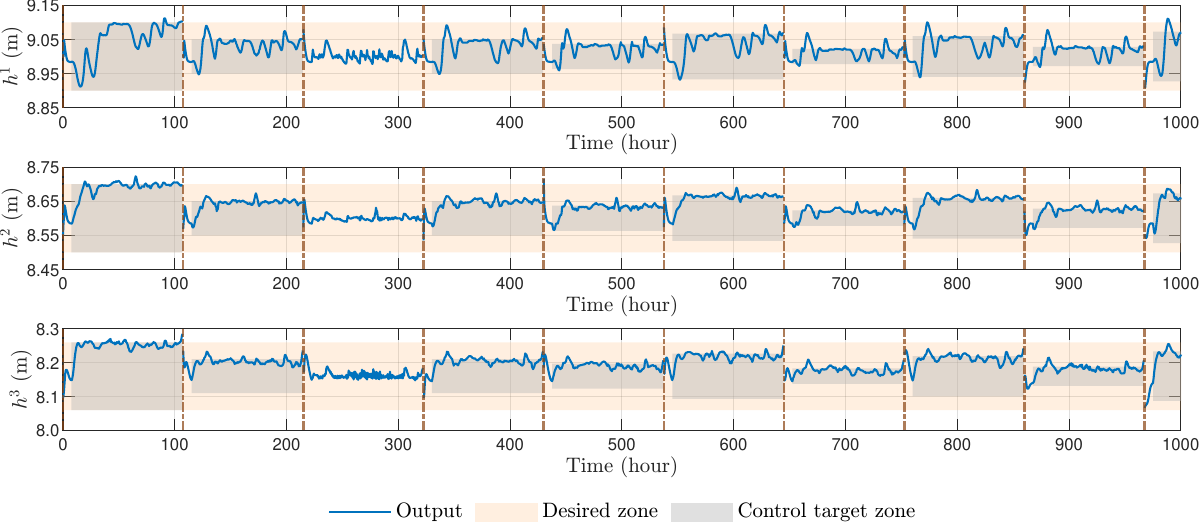}
  \caption{Concatenated trajectories of multiple BO evaluation steps.}
  \label{fig:bo_opt_traj}
\end{figure}

In each BO evaluation step, an initial trajectory of $T_{ini}$ steps is generated using a PID controller~\citep{astrom1995PID} to initialize the DeePC controller. The economic zone DeePC controller then performs closed-loop control for $T_{b}$ steps. After each evaluation step, the objective function value is computed following~\eqref{eqn:bo_obj}, and the Gaussian process (GP) surrogate model of the objective function is updated before selecting the next candidate point $\alpha$. The process is repeated for $W_{\max}$ times. Figure~\ref{fig:bo_opt_traj} illustrates the BO process by sequentially concatenating the output trajectories from each evaluation step. The brown vertical lines mark the beginning of each evaluation, and each segment in the plot represents the system output under a specific candidate parameter setting.

\begin{figure}[t!]
  \centering
  \subfloat[Evolution of the acquisition function value in the BO process, with colors denoting evaluation steps.]{\label{fig:bo_opt_single_acq}\includegraphics[width=0.47\textwidth]{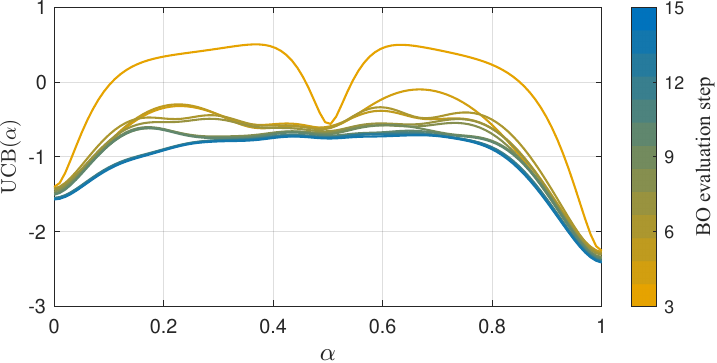}} \quad 
  \subfloat[GP surrogate model after 16 evaluation steps, with the red marker showing the optimized parameter $\alpha^{*}$.]{\label{fig:bo_opt_single}\includegraphics[width=0.47\textwidth]{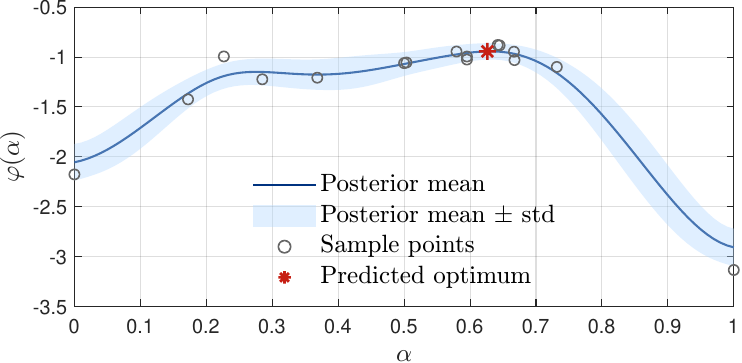}}
  \caption{BO process and the resulting GP surrogate model. The optimization process efficiently explores the parameter space and converges to the region with the highest predicted performance.}
  \label{fig:bo_opt}
\end{figure}

Figure~\ref{fig:bo_opt} illustrates the evolution of the acquisition function through the optimization process and the resulting GP surrogate model after optimization. As shown in Figure \ref{fig:bo_opt}\subref{fig:bo_opt_single_acq}, the acquisition function in the form of \eqref{eqn:acq_func} has relatively large values at the beginning. As the number of evaluation steps increases, the acquisition curve flattens, which indicates reduced uncertainty and convergence of the optimization process. 
Figure~\ref{fig:bo_opt}\subref{fig:bo_opt_single} presents the sampled points and the final GP surrogate model after all evaluation steps. The blue solid line in Figure~\ref{fig:bo_opt}\subref{fig:bo_opt_single} shows the posterior mean, and the shaded area shows the model uncertainty. The gray markers indicate the sampled points, while the red marker highlights the point with maximum posterior mean. Due to the presence of observation noise, the variance at the sampled points does not reduce to zero. The resulting GP posterior mean aligns with intuition: the objective function value is high when the scaling factor $\alpha$ is either very small or very large, which reflects poor performance due to high energy consumption in the former case and large zone tracking error in the latter.

\begin{figure}[t!]
  \centering
  \includegraphics[width=0.6\textwidth]{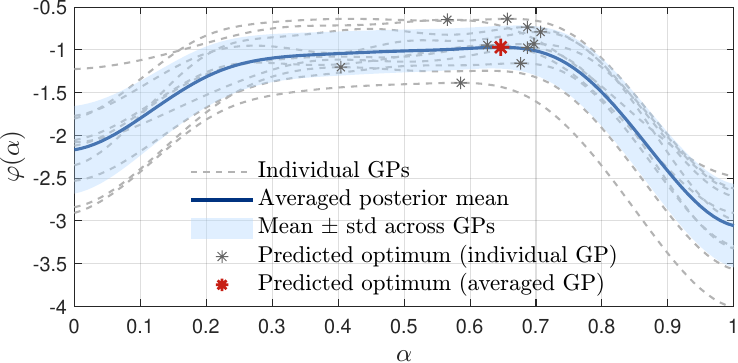}
  \caption{GP surrogate models obtained under different disturbance trajectories.}
  \label{fig:bo_opt_all}
\end{figure}

To ensure the generalizability of the optimized control target zone, BO is performed independently across 10 different disturbance trajectories. Figure~\ref{fig:bo_opt_all} presents multiple GP surrogate models fitted under different disturbance trajectories. In Figure~\ref{fig:bo_opt_all}, the dashed lines represent the posterior means of individual GP models after $W_{\max}$ evaluation steps, while the solid line represents the average of all GP posterior means. The shaded region reflects the standard deviation across the individual GP posterior means. The gray markers denote the points with the maximum posterior mean for each GP model, and the red marker highlights the point with the largest average GP mean.
As shown in Figure~\ref{fig:bo_opt_all}, the optimization results are highly consistent across disturbance trajectories. Most trained GP models share similar shapes, differing only by minor vertical shifts. Additionally, the GP models have relatively flat plateaus in the central regions, which suggests a broad range of near-optimal values. 
To consolidate the outcomes from multiple BO runs, the posterior means of all the GP models are averaged. The point with the largest averaged GP posterior mean, $\alpha^{*}=0.63$, is selected to construct the optimal control target zone following \eqref{eqn:control_target_zone}.

\subsection{Control results}
During online implementation, the optimized parameter $\alpha^{*}$ is applied for determining the control target zone $\mathbb{Z}_{t}$ for the economic zone DeePC controller (referred to as EZ-DeePC). The controller parameters are the same as those for Bayesian optimization. In the control process, an initial trajectory of $T_{ini}$ steps is generated using a PID controller~\citep{astrom1995PID}. We evaluate the closed-loop system operation performance over $N=1000$ sampling periods. 
The mean absolute error (MAE) for zone tracking is computed by averaging the absolute deviation of water levels from the desired zone:
\begin{equation}
  \text{MAE} = \frac{1}{N}\sum_{k=T_{ini}}^{N+T_{ini}-1} \min_{y^{z}_{k} \in \mathbb{Z}_{d}}\| y_{k} - y^{z}_{k} \|_{1}
\end{equation}

\begin{figure}[t!]
  \centering
  \includegraphics[width=1.0\textwidth]{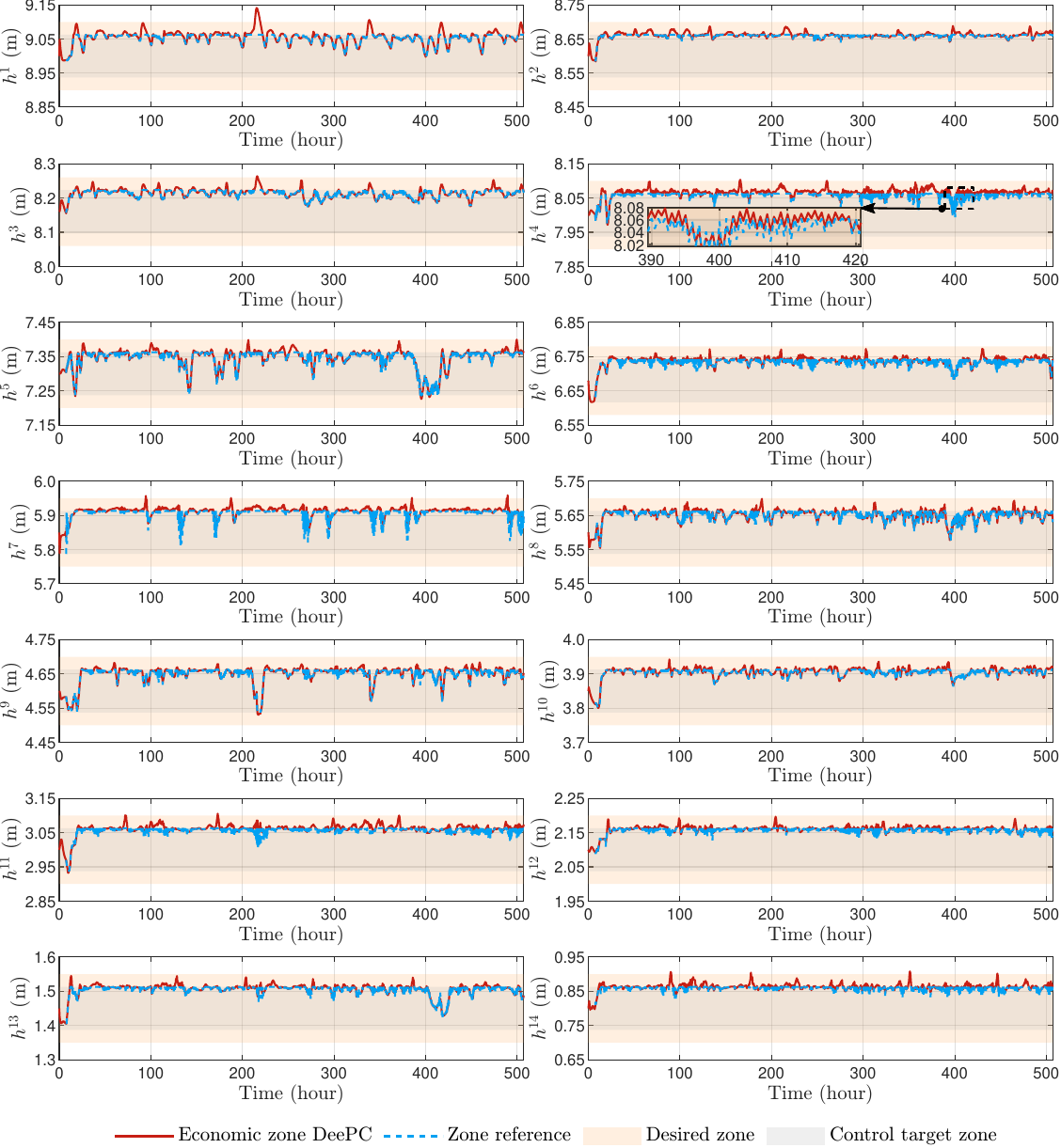}
  \caption{Output trajectories generated by EZ-DeePC controller, based on the optimized control target zone.}
  \label{fig:y_bo_res}
\end{figure}

Figure \ref{fig:y_bo_res} shows the output trajectories and the corresponding zone reference trajectories generated by the proposed EZ-DeePC controller, based on the optimized control target zone. As shown in Figure \ref{fig:y_bo_res}, the output trajectories remain within the output constraint set $\mathbb{Y}$ such that the safety constraints are satisfied. Moreover, all water levels remain within the desired zone for 97.04\% of the total time instants, which highlights the robustness of the proposed method against external disturbances. The maximum deviation from the desired zone across all branches during the control process is 0.041 m, and the zone tracking MAE is $3.73\times 10^{-4}$ m, which demonstrates good zone tracking performance. 
Although some fluctuations are observed, the water levels of all 14 branches tend to stabilize near the upper bounds of their respective control target zones. This behavior may be attributed to minimizing energy consumption by maximizing water storage and reducing total discharge from the system. 
The real-time energy consumption under the EZ-DeePC controller is shown by the solid red line in Figure~\ref{fig:energy_bo_compare}. The average energy consumption per sampling period (0.5 hour) is 33.5 kWh.

\begin{figure}[t!]
  \centering
  \includegraphics[width=1.0\textwidth]{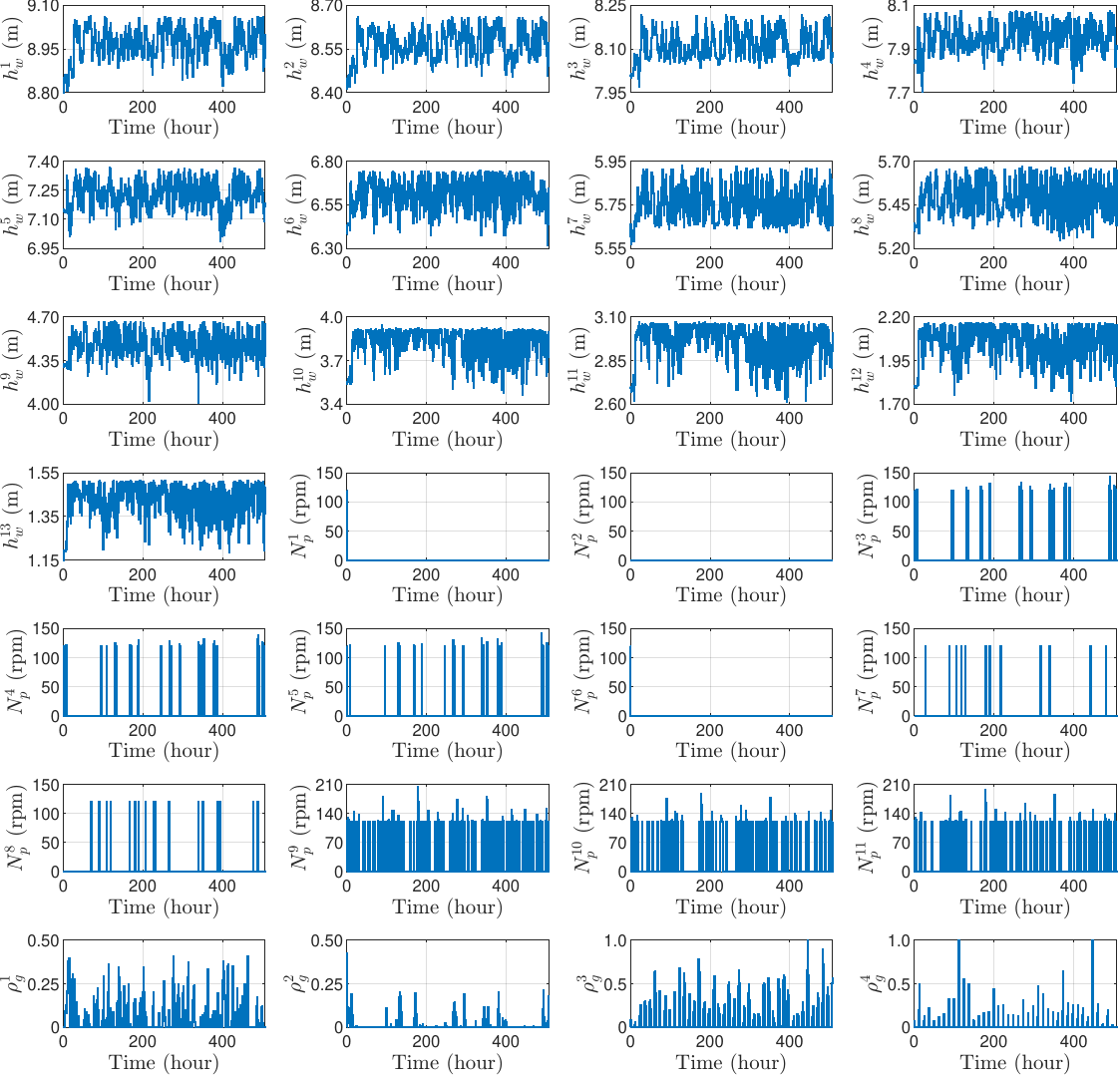}
  \caption{Control input trajectories generated by EZ-DeePC controller, based on the optimized control target zone.}
  \label{fig:u_bo_res}
\end{figure}

\begin{figure}[t!]
  \centering
  \includegraphics[width=1.0\textwidth]{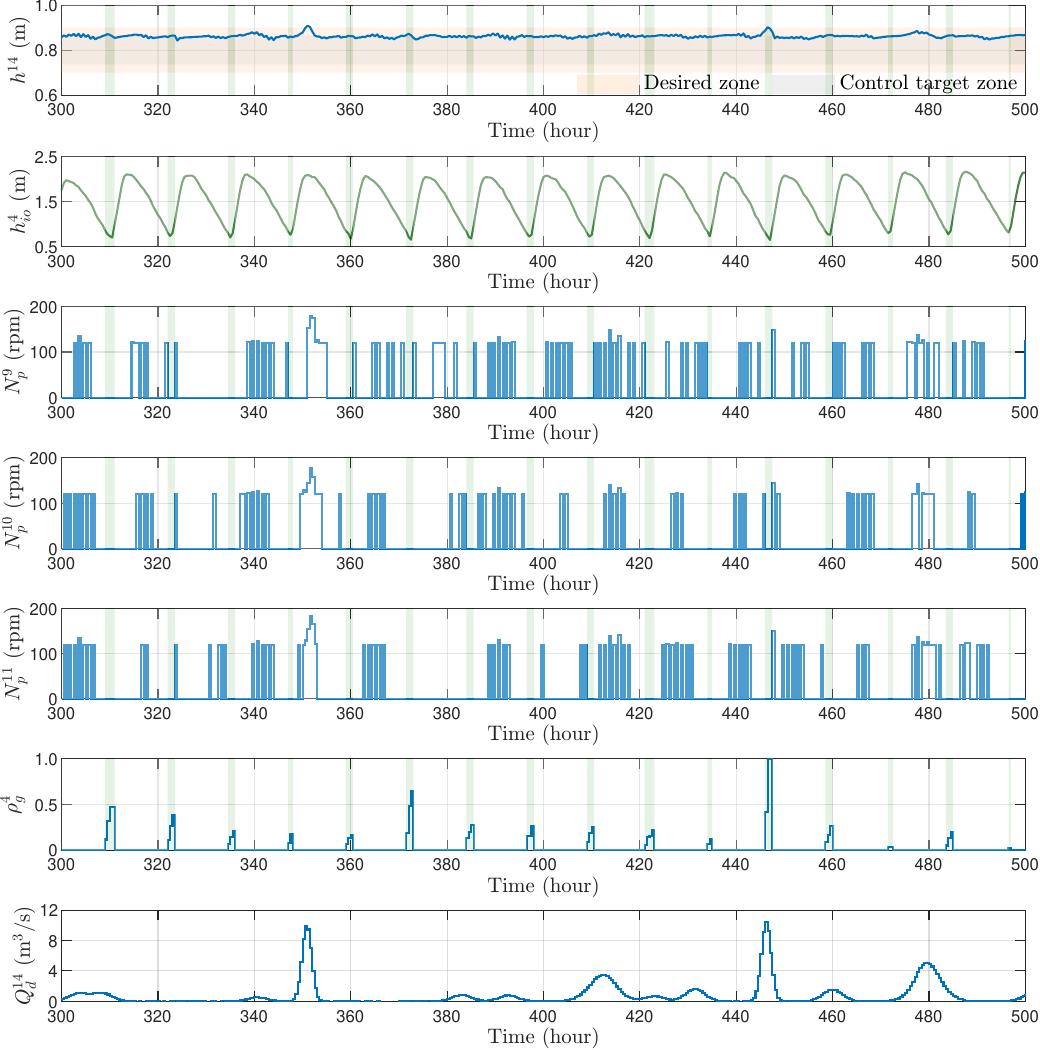}
  \caption{Output, control input, and disturbance trajectories of the 14th branch. The green area indicates the time instances when the external water level $h^{4}_{io}$ is lower than the water level of the 14th branch $h^{14}$.}
  \label{fig:station_bo_res}
\end{figure}

The control input trajectories generated by the EZ-DeePC are shown in Figure \ref{fig:u_bo_res}. Among the three types of inputs, the weir and gate inputs are frequently adjusted to regulate the water levels. As a result of minimizing energy consumption, most pumps remain off for the majority of the time, except those at the fourth station ($N_{p}^{9}$, $N_{p}^{10}$, and $N_{p}^{11}$). These pumps are located in the last branch of the system, where excess water must be pumped out during high tide. Figure \ref{fig:station_bo_res} illustrates the output, control input, and disturbance trajectories of the 14th branch over the time interval 300--500 hours. Due to tidal effects, the water level of the external river intersects with the control target zone periodically. When the water level of the branch is higher than that of the external river, water can flow through the sluice gate without requiring energy and the branch water level decreases. Conversely, when the branch water level falls below the external river level, excess water must be discharged using the pumps. In such cases, the branch water level gradually rises and stabilizes near the upper bound of the control target zone. The pumps are activated only when necessary; this results in an intermittent operation pattern.

\subsection{Comparison results}
In this section, the performance of the proposed EZ-DeePC controller with BO is compared with three alternative control approaches: economic set-point tracking DeePC (referred to as ES-DeePC;~\cite{coulson2019Dataenabled}), EZ-DeePC without BO, and EFD-based control~\citep{schutze2018Astlingen}.
For the ES-DeePC controller, the controller structure, parameters, and the offline data sequences are the same as those used in the proposed EZ-DeePC controller described in~\eqref{eqn:opt_prob_reg_zone} and~\eqref{eqn:opt_prob_reg_eco}. The only difference is that, in the ES-DeePC formulation, the width of the control target zone is set to zero, thereby reducing the zone tracking task to set-point tracking, which is the same as the DeePC method in~\cite{coulson2019Dataenabled}. 
Since the control input sequence that produces the same output sequence is not unique, the energy consumption can still be optimized even when the zone width is set to zero in the ES-DeePC formulation. For the EZ-DeePC controller without BO, the desired water-level zone is directly used as the control target zone in controller \eqref{eqn:opt_prob_reg_zone} and \eqref{eqn:opt_prob_reg_eco}.

The EFD-based control is a rule-based control method developed based on the equal-filling degree (EFD) principle~\citep{schutze2018Astlingen}. 
Specifically, the filling degree of each branch is calculated as
\begin{equation*}
  FD^i = (h^i - h^i_{\min})/ (h^i_{\max} - h^i_{\min})
\end{equation*}
where $h^i$ is the water level of the $i$th branch; $h^i_{\min}$ and $h^i_{\max}$ are the lower and upper bounds of the desired zone for the $i$th branch, respectively. The mean filling degree of each branch is denoted by $FD_{\text{mean}}$.
The weir height of the $i$th branch is adjusted based on the difference between the filling degree of the current branch and $FD_{\text{mean}}$:
\begin{equation*}
  h_w^i(k) = 
  \begin{cases}
    h^i_{\min}, & \text{if } FD^i > FD_{\text{mean}} + 0.25\\
    (h^i_{\max} + h^i_{\min})/2, & \text{if } FD^i < FD_{\text{mean}} - 0.25\\
    h_w^i(k-1), & \text{otherwise}
  \end{cases}
\end{equation*} 
For branches equipped with pumps and gates, control actions are triggered based on the filling degree and corresponding activation conditions. When the filling degree drops below 0.2 and the activation condition is satisfied, the inflow pump or gate is activated and remains in operation until the filling degree rises above 0.5.
Conversely, when the filling degree exceeds 0.8 and the activation condition is met, the outflow pump or gate operates until the filling degree decreases below 0.5. During operation, the shaft speed of each pump is set to 120 rpm when the water level remains within the desired zone ($FD^i \in [0,1]$), and is increased to 250 rpm once the water level exceeds the upper bound. Similarly, the gate opening ratio is maintained at 0.24 within the desired zone, and is adjusted to 0.5 when the water level exceeds the zone limit.

\begin{figure}[t!]
  \centering
  \includegraphics[width=1.0\textwidth]{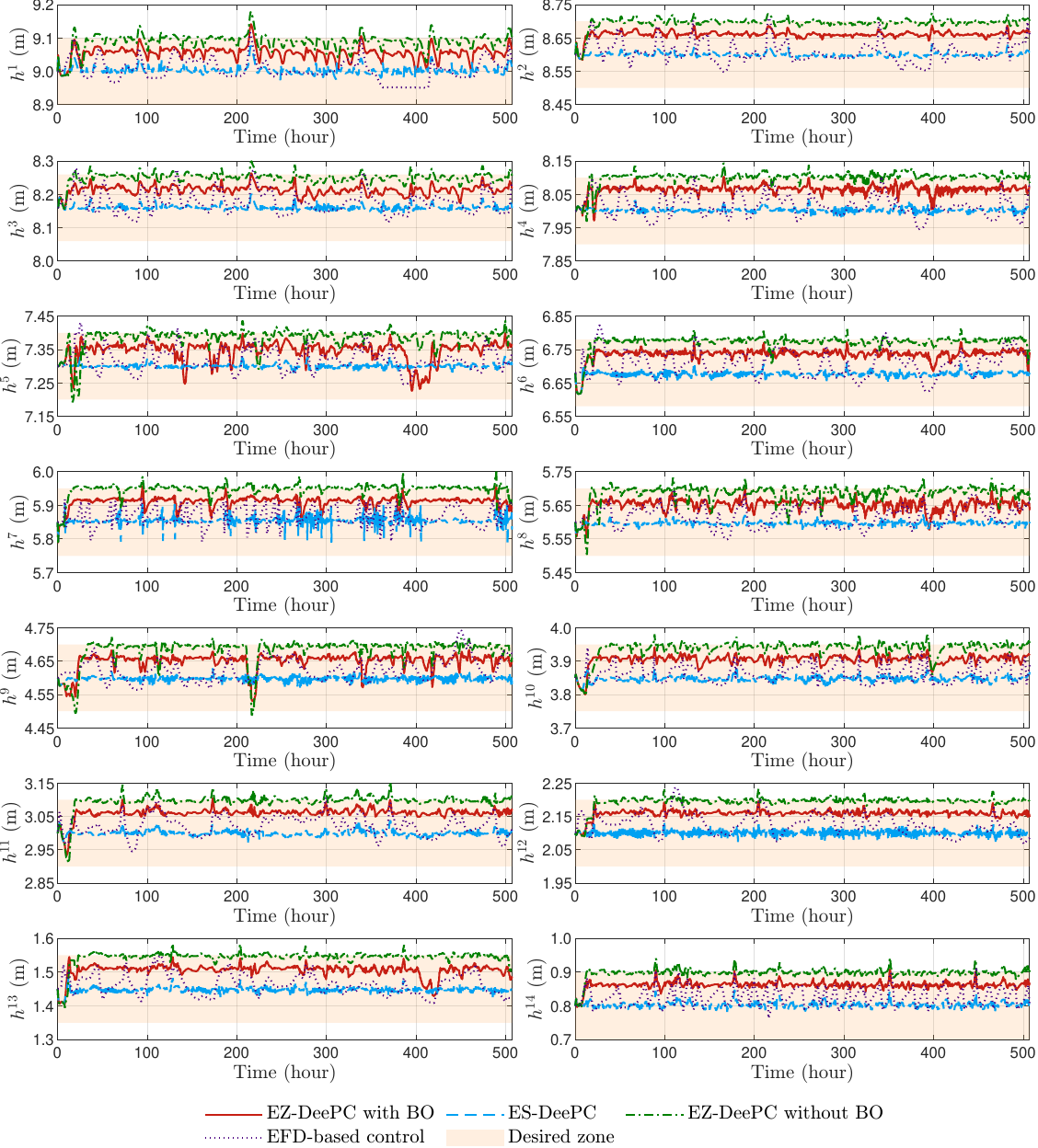}
  \caption{Output trajectories generated by EZ-DeePC with BO, ES-DeePC~\citep{coulson2019Dataenabled}, EZ-DeePC without BO, and EFD-based control~\citep{schutze2018Astlingen}.}
  \label{fig:y_bo_compare}
\end{figure}

The closed-loop output trajectories obtained from the four controllers are shown in Figure \ref{fig:y_bo_compare}.
While the ES-DeePC controller maintains the water levels near the set-point, it produces oscillatory trajectories in certain branches. For instance, the water level of the 7th branch shows noticeable fluctuations, which are observed to result from the alternating operation of the gate and pumps at the second station.
Using the EZ-DeePC without BO, the water levels exhibit a similar pattern to those of the proposed EZ-DeePC method with BO, but frequently deviate from the desired zone, which highlights the importance of zone contraction in the controller design.
Under the EFD-based control developed following~\cite{schutze2018Astlingen}, the water levels generally fluctuate within the desired zone, but occasionally exceed the zone boundaries due to disturbance inflows.

\begin{figure}[t!]
  \centering
  \includegraphics[width=1.0\textwidth]{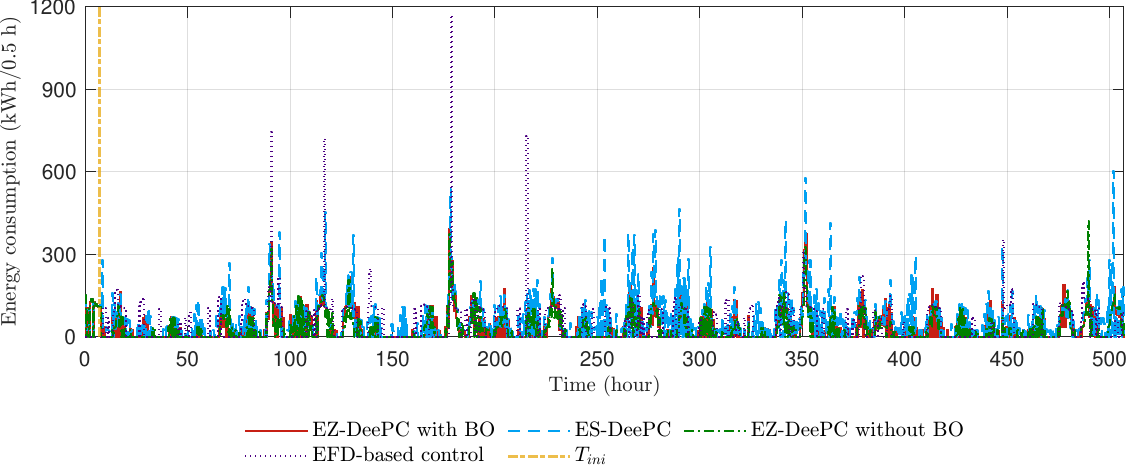}
  \caption{Comparison of energy consumption for EZ-DeePC with BO, ES-DeePC~\citep{coulson2019Dataenabled}, EZ-DeePC without BO, and EFD-based control~\citep{schutze2018Astlingen}.}
  \label{fig:energy_bo_compare}
\end{figure}

Figure~\ref{fig:energy_bo_compare} compares the real-time energy consumption profiles produced by the four controllers over the first 250 hours, where the y-axis value denotes the energy consumption over the corresponding 0.5-hour sampling period. For comparison, Table~\ref{tab:res_compare} presents the zone tracking MAE, maximum deviation from the desired zone across all branches, percentage of time instants with zone violation, and average energy consumption results for the four control methods.
The proposed method (i.e., EZ-DeePC with BO) achieves a 98.82\% reduction in zone tracking MAE, a 47.26\% reduction in maximum zone deviation, and a 96.95\% reduction in the frequency of zone violations compared with EZ-DeePC without BO; 
the corresponding reductions relative to the EFD-based control  are 83.57\% in zone tracking MAE, 4.61\% in maximum zone violation, and 74.96\% in zone violation frequency, respectively.
In the 500-hour simulation period, the proposed method maintains the water levels within the desired zone for 485 hours, which is 469 hours longer than EZ-DeePC without BO, and 44 hours longer than the EFD-based control. Assuming similar disturbance conditions throughout the year, this corresponds to approximately 342 additional days of zone maintenance compared with EZ-DeePC without BO, and 32 additional days compared with the EFD-based control.

In addition, the proposed method reduces average energy consumption by 44.08\% compared with ES-DeePC and by 22.44\% compared with EFD-based control. 
Relative to ES-DeePC, the proposed method saves $2.64\times10^{4}$~kWh energy during the 500-hour simulation period, which corresponds to an estimated annual reduction of $4.63\times10^{5}$~kWh under similar disturbance patterns.
Compared with EFD-based control, the energy saving reaches $9.69\times10^{3}$~kWh over the 500-hour simulation and approximately $1.70\times10^{5}$~kWh on an annual basis, which can be considered a substantial decrease in energy consumption and operating costs for typical water utilities.
These results indicate the superiority of the proposed method. Although ES-DeePC method can achieve zero zone tracking error, it incurs significantly higher average energy consumption compared with the proposed approach. On the other hand, while EZ-DeePC without BO yields the lowest average energy consumption, it exhibits larger zone tracking errors, primarily because the enlarged control target zone compromises the robustness of the controller against disturbances.

{
  \renewcommand{\arraystretch}{1.2}
  \begin{table}[!t]
    \centering
    \caption{Performance comparison of the proposed method, EZ-DeePC without BO, ES-DeePC~\citep{coulson2019Dataenabled}, and EFD-based control~\citep{schutze2018Astlingen}.}
    \resizebox{\textwidth}{!}{
    \begin{tabular}{ccccc}
      \toprule 
      & \makecell{Zone tracking\\ MAE (m)} & \makecell{Maximum zone\\ deviation (m)}  & \makecell{Zone violation\\ percentage (\%)} & \makecell{Average energy\\ consumption \\ (kWh/0.5 hour)} \\ \midrule 
      EZ-DeePC with BO (proposed) & $3.73\times 10^{-4}$ & $4.14\times 10^{-2}$ & 2.96 & 33.50 \\ 
      EZ-DeePC without BO & $3.16\times 10^{-2}$ &  $7.85\times 10^{-2}$ & 96.75 & 29.19 \\ 
      ES-DeePC &  0 & 0 & 0 & 59.91 \\ 
      EFD-based control & $2.27\times 10^{-3}$ &  $4.34\times 10^{-2}$ & 11.82 & 43.19 \\ 
      \bottomrule
    \end{tabular}}
    \label{tab:res_compare}
  \end{table}
}

Overall, the proposed method extends conventional DeePC~\citep{coulson2019Dataenabled} by incorporating zone-tracking and energy-minimization objectives, while it also enables adaptive tuning of the control target zone. These enhancements allow the controller to regulate water levels effectively and efficiently under varying operating conditions. Consequently, the superior performance of the proposed method, compared with baseline control designs, can be attributed to several key factors.
First, unlike ES-DeePC, which aims to maintain operation at a single water-level reference, the proposed method aims to maintain the water level within a specified zone. Incorporating zone-tracking enhances the robustness of the controller and provides greater flexibility in adjusting control actions, thereby contributing to the reduction in overall energy consumption in the connected open water system.
Second, compared to EZ-DeePC without BO, the integration of BO in the proposed method further improves robustness to external disturbances and uncertainties, while avoiding excessive energy use by adaptively selecting the optimal control target zone. This adaptive mechanism increases the applicability of the proposed approach under complex and time-varying operating conditions.
Third, in contrast to rule-based control methods such as the EFD-based control~\citep{schutze2018Astlingen}, the proposed method has predictive capability and employs an optimization-based strategy that determines control actions over a receding horizon. This proactive approach enhances both energy efficiency and water-level zone tracking performance.

Additional comparisons are conducted with data-based MPC and deep reinforcement learning (DRL) methods. 
For the data-based MPC, a deep learning-based Koopman modeling and predictive control approach~\citep{han2020Deep} (referred to as DKO-MPC), which is a representative data-driven MPC framework for nonlinear systems, is applied to the open water system. Following the existing studies in~\cite{putri2024Datadriven,zeng2025Physicsinformed}, the control objective of DKO-MPC approach~\citep{han2020Deep} is to track the center of the desired zone $y_c$. For the DRL-based controller (referred to as DRL), a state-of-the-art DRL control method, twin delayed DDPG (TD3) algorithm~\citep{fujimoto2018Addressing}, was employed. The reward function is defined as a weighted sum of the zone tracking loss and energy consumption.

The water-level zone-tracking and energy consumption metrics of the proposed method, DKO-MPC, and DRL are summarized in Table~\ref{tab:res_rl_mpc}. 
Due to mismatch between the actual system and the DKO-based surrogate model, DKO-MPC~\citep{han2020Deep} does not provide satisfactory water-level zone-tracking performance for the connected open water system. Moreover, since energy efficiency is not considered in the optimization process, the energy consumption of DKO-MPC is considerably higher that of the proposed method. 
For the DRL controller based on~\cite{fujimoto2018Addressing}, the high dimensionality of the action space hinders the learning of an effective control policy, which results in substantially larger zone-tracking errors and higher energy consumption compared with the proposed method. In addition, system constraints cannot be explicitly enforced during DRL training, which may lead to unsafe operating conditions.

{
  \renewcommand{\arraystretch}{1.2}
  \begin{table}[!t]
    \centering
    \caption{Performance comparison of the proposed method, DKO-MPC~\citep{han2020Deep}, and DRL~\citep{fujimoto2018Addressing}.}
    \resizebox{\textwidth}{!}{
    \begin{tabular}{ccccc}
      \toprule
      & \makecell{Zone tracking\\ MAE (m)} & \makecell{Maximum zone\\ deviation (m)}  & \makecell{Zone violation\\ percentage (\%)} & \makecell{Average energy\\ consumption \\(kWh/0.5 hour)} \\ 
      \midrule
      EZ-DeePC with BO & $3.73\times 10^{-4}$ & $4.14\times 10^{-2}$ & 2.96 & 33.50 \\ 
      DKO-MPC & $7.63\times 10^{-2}$ & $2.28\times 10^{-1}$ & 93.10 & 302.81 \\
      DRL & $2.50\times 10^{-1}$ &  $2.37\times 10^{-1}$ & 99.61 & 72.28 \\ 
      \bottomrule
    \end{tabular}}
    \label{tab:res_rl_mpc}
  \end{table}
  }

\subsection{Robustness validation under varied system configurations}
To further evaluate the robustness and generalizability of the proposed economic zone DeePC control approach, additional simulations are conducted under modified open water system configurations.
Specifically, we consider the following five configurations, denoted as $C_1$ to $C_5$.
\begin{itemize}
  \item Configurations $C_1$ and $C_2$ represent systems with varied backwater areas. In these two configurations, the backwater areas of the 14 branches, denoted by $A_1$ and $A_2$ (in m$^2$), are randomly sampled from the interval $[0.75 A_{0}, 1.25A_{0}]$, where $A_0$ represents the backwater areas of the 14 branches of the original system. The resulting settings are:
  \begin{equation*}
    \begin{aligned}
    A_{1} = &[155046,  22987,  37676,  40663,  56653,  66828, 273796, 46858,  97721, 441777,  83219,\\
    &198660, 157173, 958163]\\
    \end{aligned}
  \end{equation*}
  \begin{equation*}
    \begin{aligned}
    A_{2} = &[126374, 31919,  45440,  34306,  48924,  91276, 214056, 55456,  85720, 462737, 101905,\\ 
    &189758, 145026, 719617]
    \end{aligned}
  \end{equation*}
  where $A_1$ and $A_2$ represent the backwater areas of the 14 branches for configurations $C_1$ and $C_2$, respectively.
  \item Configurations $C_3$ and $C_4$ represent systems with different disturbance inflow patterns. As shown in Figure~\ref{fig:dist_compare}, the disturbance inflow in $C_3$ is intermittent with higher peaks, reflecting short but intense rainfall events, while the disturbance inflow in $C_4$ exhibits longer duration and lower intensity, corresponding to sustained mild rainfall episodes.
  \begin{figure}[!t]
    \centering
    \includegraphics[width=0.9\textwidth]{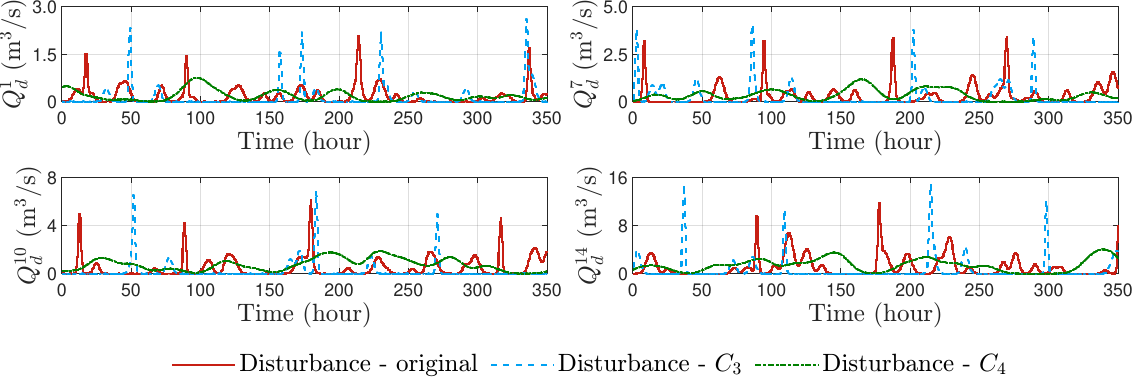}
    \caption{Comparison of disturbance patterns in system configurations $C_3$ and $C_4$.}
    \label{fig:dist_compare}
  \end{figure}
  \item Configuration $C_5$ differs from the original system by including an additional branch and a weir. The new branch is connected between the 8th branch and the 9th branch, and a weir is placed immediately downstream of the branch. The backwater area of the new branch is set to 71942 m$^{2}$; the center of the desired zone of the branch is set to 5.1 m. The operational input constraint for the added weir is specified as $h_{w} \in [2.7\ \text{m}, 6.0\ \text{m}]$. 
\end{itemize}
The other settings of the five additional configurations are the same as the original connected open water system. 

For each configuration, the proposed control method is compared with the EFD-based control developed following~\cite{schutze2018Astlingen}, which serves as the baseline. The water-level zone-tracking performance and energy consumption of the evaluations are summarized in Table~\ref{tab:robustness}. 
The reduced zone violation percentages and average energy consumption in Table~\ref{tab:robustness} demonstrate that the proposed method outperforms EFD-based control in both zone-tracking performance and energy efficiency for all five configurations. These additional evaluations further demonstrate the robustness and generalizability of the proposed method, as well as its superior performance over EFD-based control across varying system parameters, disturbances, and configurations.

Owing to the fully data-driven formulation, the proposed method eliminates the need for explicit process modeling. This feature allows the controller to generalize across different system configurations and disturbance patterns while maintaining good water-level zone-tracking performance and energy efficiency.

{
  \renewcommand{\arraystretch}{1.2}
  \begin{table}[t]
    \centering
    \caption{Performance comparison under varied system configurations.}
    \resizebox{\textwidth}{!}{
    \begin{tabular}{cccccc}
      \toprule
      Config. & Method & \makecell{Zone tracking\\ MAE (m)} & \makecell{Maximum zone\\ deviation (m)}  & \makecell{Zone violation\\ percentage (\%)} & \makecell{Average energy\\ consumption \\(kWh/0.5 hour)} \\ 
        \midrule
      \multirow{2}{*}{$C_1$} & EZ-DeePC with BO & $7.39\times 10^{-5}$ & $1.60\times 10^{-2}$ & 1.77 & 18.00 \\
                          & EFD-based control & $6.67\times 10^{-3}$ & $9.88\times 10^{-2}$ & 16.45 & 36.15 \\ \cmidrule(lr){1-6}
      \multirow{2}{*}{$C_2$} & EZ-DeePC with BO & $1.30\times 10^{-3}$ & $5.48\times 10^{-2}$ & 8.77 & 25.31 \\
                          & EFD-based control & $1.80\times 10^{-3}$ & $9.34\times 10^{-2}$ & 44.24 & 52.76 \\ \cmidrule(lr){1-6}
      \multirow{2}{*}{$C_3$} & EZ-DeePC with BO & $2.78\times 10^{-4}$ & $2.82\times 10^{-2}$ & 3.84 & 12.71 \\
                          & EFD-based control & $1.57\times 10^{-3}$ & $5.87\times 10^{-2}$ & 8.27 & 19.47 \\ \cmidrule(lr){1-6}
      \multirow{2}{*}{$C_4$} & EZ-DeePC with BO & 0 & 0 & 0 & 31.58 \\
                          & EFD-based control & $8.84\times 10^{-3}$ & $3.58\times 10^{-2}$ & 37.14 & 51.19 \\
                          \cmidrule(lr){1-6}
      \multirow{2}{*}{$C_5$} & EZ-DeePC with BO & $1.81\times 10^{-4}$ & $2.24\times 10^{-2}$ & 3.25 & 19.47 \\
                          & EFD-based control & $3.28\times 10^{-3}$ & $5.08\times 10^{-2}$ & 15.07 & 35.03 \\ 
      \bottomrule
    \end{tabular}}
    \label{tab:robustness}
  \end{table}
}

\section{Conclusion}
In this paper, we proposed a data-driven mixed-integer economic zone predictive control approach to regulate the water levels and minimize the energy consumption of a connected open water system. The controller was designed purely from system input and output data, and its formulation is consistent with the operational objectives of ensuring safe and energy-efficient operation. Lexicographic optimization was employed to simultaneously handle the two control objectives. Bayesian optimization was conducted to determine an appropriate control target zone for the proposed controller; this way, mismatches induced by system nonlinearity and disturbances are appropriately addressed.
Extensive simulations and comparative analyses were conducted, demonstrating that the proposed method can simultaneously regulate water levels across all branches and reduce system energy consumption. Specifically, the water levels of the 14 branches are maintained within the desired zone for 97.04\% of the operating time, with an average energy consumption of 33.5 kWh per 0.5 hour. The proposed method reduces the zone tracking MAE by 98.82\% compared to economic zone DeePC without BO-based control target zone identification, and reduces energy consumption by 44.08\% compared to economic set-point tracking DeePC. Additionally, it significantly outperforms the EFD-based in both zone tracking performance and energy efficiency. 

Despite the promising results, this study has several limitations that should be addressed in future work. The evaluations were conducted solely through simulations, and the simplified simulator may not fully capture the complex hydraulic dynamics and spatial variability of disturbances in the connected open water systems.
Future research will focus on developing a Saint--Venant equation-based simulation platform to more accurately and comprehensively capture the hydrodynamic behavior and spatially distributed disturbances of real open water systems.
Moreover, the proposed data-driven economic zone control framework will be further enhanced by incorporating disturbance forecasting and time-delay effects, and extended to high-fidelity simulators and lab-scale experimental setups to evaluate its effectiveness and applicability.

\section*{Acknowledgment}
This research is supported by the National Research Foundation, Singapore, and PUB, Singapore's National Water Agency under its RIE2025 Urban Solutions and Sustainability (USS) (Water) Centre of Excellence (CoE) Programme, awarded to Nanyang Environment \& Water Research Institute (NEWRI), Nanyang Technological
University, Singapore (NTU). This research is also supported by the Ministry of Education, Singapore, under its Academic Research Fund Tier 1 (RG95/24 and RG63/22). Any opinions, findings and conclusions or recommendations expressed in this material are those of the author(s) and do not reflect the views of the National Research Foundation, Singapore, and PUB, Singapore's National Water Agency.


\begin{thebibliography}{65}
\expandafter\ifx\csname natexlab\endcsname\relax\def\natexlab#1{#1}\fi
\providecommand{\url}[1]{\texttt{#1}}
\providecommand{\href}[2]{#2}
\providecommand{\path}[1]{#1}
\providecommand{\DOIprefix}{doi:}
\providecommand{\ArXivprefix}{arXiv:}
\providecommand{\URLprefix}{URL: }
\providecommand{\Pubmedprefix}{pmid:}
\providecommand{\doi}[1]{\href{http://dx.doi.org/#1}{\path{#1}}}
\providecommand{\Pubmed}[1]{\href{pmid:#1}{\path{#1}}}
\providecommand{\bibinfo}[2]{#2}
\ifx\xfnm\relax \def\xfnm[#1]{\unskip,\space#1}\fi
\bibitem[{Anilkumar et~al.(2016)Anilkumar, Padhiyar and Moudgalya}]{anilkumar2016Lexicographic}
\bibinfo{author}{Anilkumar, M.}, \bibinfo{author}{Padhiyar, N.}, \bibinfo{author}{Moudgalya, K.}, \bibinfo{year}{2016}.
\newblock \bibinfo{title}{Lexicographic optimization based {{MPC}}: Simulation and experimental study}.
\newblock \bibinfo{journal}{Comput. Chem. Eng.} \bibinfo{volume}{88}, \bibinfo{pages}{135--144}.
\newblock \DOIprefix\doi{10.1016/j.compchemeng.2016.02.002}.
\bibitem[{{\AA}str{\"o}m and H{\"a}gglund(1995)}]{astrom1995PID}
\bibinfo{author}{{\AA}str{\"o}m, K.J.}, \bibinfo{author}{H{\"a}gglund, T.}, \bibinfo{year}{1995}.
\newblock \bibinfo{title}{{{PID}} Controllers: Theory, Design, and Tuning}.
\newblock \bibinfo{publisher}{{International Society for Measurement and Control}}, \bibinfo{address}{Research Triangle Park, N.C}.
\bibitem[{Balla et~al.(2022)Balla, Bendtsen, Schou, Kalles{\o}e and {Ocampo-Martinez}}]{balla2022Learningbased}
\bibinfo{author}{Balla, K.M.}, \bibinfo{author}{Bendtsen, J.D.}, \bibinfo{author}{Schou, C.}, \bibinfo{author}{Kalles{\o}e, C.S.}, \bibinfo{author}{{Ocampo-Martinez}, C.}, \bibinfo{year}{2022}.
\newblock \bibinfo{title}{A learning-based approach towards the data-driven predictive control of combined wastewater networks -- an experimental study}.
\newblock \bibinfo{journal}{Water Res.} \bibinfo{volume}{221}, \bibinfo{pages}{118782}.
\newblock \DOIprefix\doi{10.1016/j.watres.2022.118782}.
\bibitem[{Becker et~al.(2024)Becker, Jagtenberg, Horv{\'a}th, Mitchell and {Rodr{\'i}guez-Sarasty}}]{becker2024Optimization}
\bibinfo{author}{Becker, B.P.J.}, \bibinfo{author}{Jagtenberg, C.J.}, \bibinfo{author}{Horv{\'a}th, K.}, \bibinfo{author}{Mitchell, A.}, \bibinfo{author}{{Rodr{\'i}guez-Sarasty}, J.A.}, \bibinfo{year}{2024}.
\newblock \bibinfo{title}{Optimization methods in water system operation}.
\newblock \bibinfo{journal}{WIREs Water} \bibinfo{volume}{11}, \bibinfo{pages}{e1756}.
\newblock \DOIprefix\doi{10.1002/wat2.1756}.
\bibitem[{Berberich et~al.(2021)Berberich, K{\"o}hler, M{\"u}ller and Allg{\"o}wer}]{berberich2021Datadriven}
\bibinfo{author}{Berberich, J.}, \bibinfo{author}{K{\"o}hler, J.}, \bibinfo{author}{M{\"u}ller, M.A.}, \bibinfo{author}{Allg{\"o}wer, F.}, \bibinfo{year}{2021}.
\newblock \bibinfo{title}{Data-driven model predictive control with stability and robustness guarantees}.
\newblock \bibinfo{journal}{IEEE Trans. Autom. Control} \bibinfo{volume}{66}, \bibinfo{pages}{1702--1717}.
\newblock \DOIprefix\doi{10.1109/TAC.2020.3000182}.
\bibitem[{Bos(1989)}]{bos1989Discharge}
\bibinfo{editor}{Bos, M.G.} (Ed.), \bibinfo{year}{1989}.
\newblock \bibinfo{title}{Discharge Measurement Structures}.
\newblock \bibinfo{publisher}{{International Institute for Land Reclamation and Improvement}}, \bibinfo{address}{Wageningen}.
\bibitem[{Breckpot et~al.(2013)Breckpot, Agudelo and De~Moor}]{breckpot2013Flood}
\bibinfo{author}{Breckpot, M.}, \bibinfo{author}{Agudelo, O.M.}, \bibinfo{author}{De~Moor, B.}, \bibinfo{year}{2013}.
\newblock \bibinfo{title}{Flood control with model predictive control for river systems with water reservoirs}.
\newblock \bibinfo{journal}{J. Irrig. Drain. Eng.} \bibinfo{volume}{139}, \bibinfo{pages}{532--541}.
\newblock \DOIprefix\doi{10.1061/(ASCE)IR.1943-4774.0000577}.
\bibitem[{Breschi et~al.(2023a)Breschi, Chiuso and Formentin}]{breschi2023Datadriven}
\bibinfo{author}{Breschi, V.}, \bibinfo{author}{Chiuso, A.}, \bibinfo{author}{Formentin, S.}, \bibinfo{year}{2023}a.
\newblock \bibinfo{title}{Data-driven predictive control in a stochastic setting: A unified framework}.
\newblock \bibinfo{journal}{Automatica J. IFAC} \bibinfo{volume}{152}, \bibinfo{pages}{110961}.
\newblock \DOIprefix\doi{10.1016/j.automatica.2023.110961}.
\bibitem[{Breschi et~al.(2023b)Breschi, Fabris, Formentin and Chiuso}]{breschi2023Uncertaintyaware}
\bibinfo{author}{Breschi, V.}, \bibinfo{author}{Fabris, M.}, \bibinfo{author}{Formentin, S.}, \bibinfo{author}{Chiuso, A.}, \bibinfo{year}{2023}b.
\newblock \bibinfo{title}{Uncertainty-aware data-driven predictive control in a stochastic setting}.
\newblock \bibinfo{journal}{IFAC-Pap.} \bibinfo{volume}{56}, \bibinfo{pages}{10083--10088}.
\newblock \DOIprefix\doi{10.1016/j.ifacol.2023.10.878}.
\bibitem[{Brochu et~al.(2010)Brochu, Cora and de~Freitas}]{brochu2010Tutorial}
\bibinfo{author}{Brochu, E.}, \bibinfo{author}{Cora, V.M.}, \bibinfo{author}{de~Freitas, N.}, \bibinfo{year}{2010}.
\newblock \bibinfo{title}{A tutorial on {{Bayesian}} optimization of expensive cost functions, with application to active user modeling and hierarchical reinforcement learning}.
\newblock \DOIprefix\doi{10.48550/arXiv.1012.2599}, \href{http://arxiv.org/abs/1012.2599}{{\tt arXiv:1012.2599}}.
\bibitem[{Byrd et~al.(2006)Byrd, Nocedal and Waltz}]{byrd2006Knitro}
\bibinfo{author}{Byrd, R.H.}, \bibinfo{author}{Nocedal, J.}, \bibinfo{author}{Waltz, R.A.}, \bibinfo{year}{2006}.
\newblock \bibinfo{title}{Knitro: An integrated package for nonlinear optimization}, in: \bibinfo{editor}{Di~Pillo, G.}, \bibinfo{editor}{Roma, M.} (Eds.), \bibinfo{booktitle}{Large-Scale Nonlinear Optimization}. \bibinfo{publisher}{Springer US}, \bibinfo{address}{Boston, MA}, pp. \bibinfo{pages}{35--59}.
\newblock \DOIprefix\doi{10.1007/0-387-30065-1_4}.
\bibitem[{Castelletti et~al.(2023)Castelletti, Ficch{\`i}, Cominola, Segovia, Giuliani, Wu, Lucia, {Ocampo-Martinez}, De~Schutter and Maestre}]{castelletti2023Model}
\bibinfo{author}{Castelletti, A.}, \bibinfo{author}{Ficch{\`i}, A.}, \bibinfo{author}{Cominola, A.}, \bibinfo{author}{Segovia, P.}, \bibinfo{author}{Giuliani, M.}, \bibinfo{author}{Wu, W.}, \bibinfo{author}{Lucia, S.}, \bibinfo{author}{{Ocampo-Martinez}, C.}, \bibinfo{author}{De~Schutter, B.}, \bibinfo{author}{Maestre, J.M.}, \bibinfo{year}{2023}.
\newblock \bibinfo{title}{Model predictive control of water resources systems: A review and research agenda}.
\newblock \bibinfo{journal}{Annu. Rev. Control} \bibinfo{volume}{55}, \bibinfo{pages}{442--465}.
\newblock \DOIprefix\doi{10.1016/j.arcontrol.2023.03.013}.
\bibitem[{Christofides et~al.(2013)Christofides, Scattolini, {Mu{\~n}oz de la Pe{\~n}a} and Liu}]{christofides2013Distributed}
\bibinfo{author}{Christofides, P.D.}, \bibinfo{author}{Scattolini, R.}, \bibinfo{author}{{Mu{\~n}oz de la Pe{\~n}a}, D.}, \bibinfo{author}{Liu, J.}, \bibinfo{year}{2013}.
\newblock \bibinfo{title}{Distributed model predictive control: A tutorial review and future research directions}.
\newblock \bibinfo{journal}{Comput. Chem. Eng.} \bibinfo{volume}{51}, \bibinfo{pages}{21--41}.
\newblock \DOIprefix\doi{10.1016/j.compchemeng.2012.05.011}.
\bibitem[{Coulson et~al.(2019)Coulson, Lygeros and D{\"o}rfler}]{coulson2019Dataenabled}
\bibinfo{author}{Coulson, J.}, \bibinfo{author}{Lygeros, J.}, \bibinfo{author}{D{\"o}rfler, F.}, \bibinfo{year}{2019}.
\newblock \bibinfo{title}{Data-enabled predictive control: In the shallows of the {{DeePC}}}, in: \bibinfo{booktitle}{Eur. {{Control Conf}}.}, \bibinfo{address}{Naples, Italy}. pp. \bibinfo{pages}{307--312}.
\newblock \DOIprefix\doi{10.23919/ECC.2019.8795639}.
\bibitem[{D{\"o}rfler et~al.(2023)D{\"o}rfler, Coulson and Markovsky}]{dorfler2023Bridging}
\bibinfo{author}{D{\"o}rfler, F.}, \bibinfo{author}{Coulson, J.}, \bibinfo{author}{Markovsky, I.}, \bibinfo{year}{2023}.
\newblock \bibinfo{title}{Bridging direct and indirect data-driven control formulations via regularizations and relaxations}.
\newblock \bibinfo{journal}{IEEE Trans. Autom. Control} \bibinfo{volume}{68}, \bibinfo{pages}{883--897}.
\newblock \DOIprefix\doi{10.1109/TAC.2022.3148374}.
\bibitem[{Ferramosca et~al.(2010)Ferramosca, Limon, Gonz{\'a}lez, Odloak and Camacho}]{ferramosca2010MPC}
\bibinfo{author}{Ferramosca, A.}, \bibinfo{author}{Limon, D.}, \bibinfo{author}{Gonz{\'a}lez, A.H.}, \bibinfo{author}{Odloak, D.}, \bibinfo{author}{Camacho, E.F.}, \bibinfo{year}{2010}.
\newblock \bibinfo{title}{{{MPC}} for tracking zone regions}.
\newblock \bibinfo{journal}{J. Process Control} \bibinfo{volume}{20}, \bibinfo{pages}{506--516}.
\newblock \DOIprefix\doi{10.1016/j.jprocont.2010.02.005}.
\bibitem[{Floudas(1995)}]{floudas1995Nonlinear}
\bibinfo{author}{Floudas, C.A.}, \bibinfo{year}{1995}.
\newblock \bibinfo{title}{Nonlinear and Mixed-Integer Optimization: Fundamentals and Applications}.
\newblock \bibinfo{publisher}{Oxford University Press}.
\bibitem[{Frazier(2018)}]{frazier2018Tutorial}
\bibinfo{author}{Frazier, P.I.}, \bibinfo{year}{2018}.
\newblock \bibinfo{title}{A tutorial on {{Bayesian}} optimization}.
\newblock \DOIprefix\doi{10.48550/arXiv.1807.02811}, \href{http://arxiv.org/abs/1807.02811}{{\tt arXiv:1807.02811}}.
\bibitem[{Fujimoto et~al.(2018)Fujimoto, Hoof and Meger}]{fujimoto2018Addressing}
\bibinfo{author}{Fujimoto, S.}, \bibinfo{author}{Hoof, H.}, \bibinfo{author}{Meger, D.}, \bibinfo{year}{2018}.
\newblock \bibinfo{title}{Addressing function approximation error in actor-critic methods}, in: \bibinfo{booktitle}{Int. {{Conf}}. {{Mach}}. {{Learn}}.}, \bibinfo{publisher}{PMLR}. pp. \bibinfo{pages}{1587--1596}.
\bibitem[{Gan et~al.(2024)Gan, Jiang, Zhao, He and Duan}]{gan2024Research}
\bibinfo{author}{Gan, T.}, \bibinfo{author}{Jiang, Y.}, \bibinfo{author}{Zhao, H.}, \bibinfo{author}{He, J.}, \bibinfo{author}{Duan, H.}, \bibinfo{year}{2024}.
\newblock \bibinfo{title}{Research on low-energy consumption automatic real-time regulation of cascade gates and pumps in open-canal based on reinforcement learning}.
\newblock \bibinfo{journal}{J. Hydroinf.} \bibinfo{volume}{26}, \bibinfo{pages}{1673--1691}.
\newblock \DOIprefix\doi{10.2166/hydro.2024.020}.
\bibitem[{Golub and Loan(2013)}]{golub2013Matrix}
\bibinfo{author}{Golub, G.H.}, \bibinfo{author}{Loan, C.F.V.}, \bibinfo{year}{2013}.
\newblock \bibinfo{title}{Matrix Computations}.
\newblock \bibinfo{publisher}{JHU Press}.
\bibitem[{Gonz{\'a}lez et~al.(2009)Gonz{\'a}lez, Marchetti and Odloak}]{gonzalez2009Robust}
\bibinfo{author}{Gonz{\'a}lez, A.}, \bibinfo{author}{Marchetti, J.}, \bibinfo{author}{Odloak, D.}, \bibinfo{year}{2009}.
\newblock \bibinfo{title}{Robust model predictive control with zone control}.
\newblock \bibinfo{journal}{IET Control Theory Appl.} \bibinfo{volume}{3}, \bibinfo{pages}{121--135}.
\newblock \DOIprefix\doi{10.1049/iet-cta:20070211}.
\bibitem[{Gonz{\'a}lez and Odloak(2009)}]{gonzalez2009Stable}
\bibinfo{author}{Gonz{\'a}lez, A.H.}, \bibinfo{author}{Odloak, D.}, \bibinfo{year}{2009}.
\newblock \bibinfo{title}{A stable {{MPC}} with zone control}.
\newblock \bibinfo{journal}{J. Process Control} \bibinfo{volume}{19}, \bibinfo{pages}{110--122}.
\newblock \DOIprefix\doi{10.1016/j.jprocont.2008.01.003}.
\bibitem[{Han et~al.(2020)Han, Hao and Vaidya}]{han2020Deep}
\bibinfo{author}{Han, Y.}, \bibinfo{author}{Hao, W.}, \bibinfo{author}{Vaidya, U.}, \bibinfo{year}{2020}.
\newblock \bibinfo{title}{Deep learning of {{Koopman}} representation for control}, in: \bibinfo{booktitle}{{{IEEE Conf}}. {{Decis}}. {{Control}}}, \bibinfo{address}{Jeju, Korea (South)}. pp. \bibinfo{pages}{1890--1895}.
\newblock \DOIprefix\doi{10.1109/CDC42340.2020.9304238}.
\bibitem[{Horv{\'a}th et~al.(2019a)Horv{\'a}th, Esch, Pothof, Vreeken, Talsma and Baayen}]{horvath2019Closedloop}
\bibinfo{author}{Horv{\'a}th, K.}, \bibinfo{author}{Esch, B.V.}, \bibinfo{author}{Pothof, I.}, \bibinfo{author}{Vreeken, T.}, \bibinfo{author}{Talsma, J.}, \bibinfo{author}{Baayen, J.}, \bibinfo{year}{2019}a.
\newblock \bibinfo{title}{Closed-loop model predictive control with mixed-integer optimization of a river reach with weirs}.
\newblock \bibinfo{journal}{IFAC-Pap.} \bibinfo{volume}{52}, \bibinfo{pages}{81--87}.
\newblock \DOIprefix\doi{10.1016/j.ifacol.2019.11.013}.
\bibitem[{Horv{\'a}th et~al.(2019b)Horv{\'a}th, {van Esch}, Vreeken, Pothof and Baayen}]{horvath2019Convex}
\bibinfo{author}{Horv{\'a}th, K.}, \bibinfo{author}{{van Esch}, B.}, \bibinfo{author}{Vreeken, D.}, \bibinfo{author}{Pothof, I.}, \bibinfo{author}{Baayen, J.}, \bibinfo{year}{2019}b.
\newblock \bibinfo{title}{Convex modeling of pumps in order to optimize their energy use}.
\newblock \bibinfo{journal}{Water Resour. Res.} \bibinfo{volume}{55}, \bibinfo{pages}{2432--2445}.
\newblock \DOIprefix\doi{10.1029/2018WR023811}.
\bibitem[{Horv{\'a}th et~al.(2022)Horv{\'a}th, Van~Esch, Vreeken, Piovesan, Talsma and Pothof}]{horvath2022Potential}
\bibinfo{author}{Horv{\'a}th, K.}, \bibinfo{author}{Van~Esch, B.}, \bibinfo{author}{Vreeken, T.}, \bibinfo{author}{Piovesan, T.}, \bibinfo{author}{Talsma, J.}, \bibinfo{author}{Pothof, I.}, \bibinfo{year}{2022}.
\newblock \bibinfo{title}{Potential of model predictive control of a polder water system including pumps, weirs and gates}.
\newblock \bibinfo{journal}{J. Process Control} \bibinfo{volume}{119}, \bibinfo{pages}{128--140}.
\newblock \DOIprefix\doi{10.1016/j.jprocont.2022.10.003}.
\bibitem[{Huang et~al.(2022)Huang, Coulson, Lygeros and D{\"o}rfler}]{huang2022Decentralized}
\bibinfo{author}{Huang, L.}, \bibinfo{author}{Coulson, J.}, \bibinfo{author}{Lygeros, J.}, \bibinfo{author}{D{\"o}rfler, F.}, \bibinfo{year}{2022}.
\newblock \bibinfo{title}{Decentralized data-enabled predictive control for power system oscillation damping}.
\newblock \bibinfo{journal}{IEEE Trans. Control Syst. Technol.} \bibinfo{volume}{30}, \bibinfo{pages}{1065--1077}.
\newblock \DOIprefix\doi{10.1109/TCST.2021.3088638}.
\bibitem[{Isermann(1982)}]{isermann1982Linear}
\bibinfo{author}{Isermann, H.}, \bibinfo{year}{1982}.
\newblock \bibinfo{title}{Linear lexicographic optimization}.
\newblock \bibinfo{journal}{OR Spektrum} \bibinfo{volume}{4}, \bibinfo{pages}{223--228}.
\newblock \DOIprefix\doi{10.1007/BF01782758}.
\bibitem[{Jv et~al.(2023)Jv, Wang, Zhang, Yin and Liu}]{jv2023Lexicographic}
\bibinfo{author}{Jv, Y.}, \bibinfo{author}{Wang, Z.}, \bibinfo{author}{Zhang, Y.}, \bibinfo{author}{Yin, X.}, \bibinfo{author}{Liu, J.}, \bibinfo{year}{2023}.
\newblock \bibinfo{title}{Lexicographic optimization for economic model predictive control with zone tracking}.
\newblock \bibinfo{journal}{Chem. Eng. Res. Des.} \bibinfo{volume}{200}, \bibinfo{pages}{646--654}.
\newblock \DOIprefix\doi{10.1016/j.cherd.2023.11.041}.
\bibitem[{Kong et~al.(2023)Kong, Li, Tang, Yuan, Yang, Ji, Li and Chen}]{kong2023Predictive}
\bibinfo{author}{Kong, L.}, \bibinfo{author}{Li, Y.}, \bibinfo{author}{Tang, H.}, \bibinfo{author}{Yuan, S.}, \bibinfo{author}{Yang, Q.}, \bibinfo{author}{Ji, Q.}, \bibinfo{author}{Li, Z.}, \bibinfo{author}{Chen, R.}, \bibinfo{year}{2023}.
\newblock \bibinfo{title}{Predictive control for the operation of cascade pumping stations in water supply canal systems considering energy consumption and costs}.
\newblock \bibinfo{journal}{Appl. Energy} \bibinfo{volume}{341}, \bibinfo{pages}{121103}.
\newblock \DOIprefix\doi{10.1016/j.apenergy.2023.121103}.
\bibitem[{Li et~al.(2024)Li, Zhang, Dong, Wang, Li and Song}]{li2024Physicsaugmented}
\bibinfo{author}{Li, D.}, \bibinfo{author}{Zhang, K.}, \bibinfo{author}{Dong, H.}, \bibinfo{author}{Wang, Q.}, \bibinfo{author}{Li, Z.}, \bibinfo{author}{Song, Z.}, \bibinfo{year}{2024}.
\newblock \bibinfo{title}{Physics-augmented data-enabled predictive control for eco-driving of mixed traffic considering diverse human behaviors}.
\newblock \bibinfo{journal}{IEEE Trans. Control Syst. Technol.} \bibinfo{volume}{32}, \bibinfo{pages}{1479--1486}.
\newblock \DOIprefix\doi{10.1109/TCST.2024.3361393}.
\bibitem[{Litrico and Fromion(2009)}]{litrico2009Modeling}
\bibinfo{author}{Litrico, X.}, \bibinfo{author}{Fromion, V.}, \bibinfo{year}{2009}.
\newblock \bibinfo{title}{Modeling and Control of Hydrosystems}.
\newblock \bibinfo{publisher}{Springer London}, \bibinfo{address}{London}.
\newblock \DOIprefix\doi{10.1007/978-1-84882-624-3}.
\bibitem[{Liu and Liu(2018)}]{liu2018Economic}
\bibinfo{author}{Liu, S.}, \bibinfo{author}{Liu, J.}, \bibinfo{year}{2018}.
\newblock \bibinfo{title}{Economic model predictive control with zone tracking}.
\newblock \bibinfo{journal}{Mathematics} \bibinfo{volume}{6}, \bibinfo{pages}{65}.
\newblock \DOIprefix\doi{10.3390/math6050065}.
\bibitem[{Liu et~al.(2019)Liu, Mao and Liu}]{liu2019Modelpredictive}
\bibinfo{author}{Liu, S.}, \bibinfo{author}{Mao, Y.}, \bibinfo{author}{Liu, J.}, \bibinfo{year}{2019}.
\newblock \bibinfo{title}{Model-predictive control with generalized zone tracking}.
\newblock \bibinfo{journal}{IEEE Trans. Autom. Control} \bibinfo{volume}{64}, \bibinfo{pages}{4698--4704}.
\newblock \DOIprefix\doi{10.1109/TAC.2019.2902041}.
\bibitem[{Lu et~al.(2021)Lu, Gonz{\'a}lez, Kumar and Zavala}]{lu2021Bayesian}
\bibinfo{author}{Lu, Q.}, \bibinfo{author}{Gonz{\'a}lez, L.D.}, \bibinfo{author}{Kumar, R.}, \bibinfo{author}{Zavala, V.M.}, \bibinfo{year}{2021}.
\newblock \bibinfo{title}{Bayesian optimization with reference models: A case study in {{MPC}} for {{HVAC}} central plants}.
\newblock \bibinfo{journal}{Comput. Chem. Eng.} \bibinfo{volume}{154}, \bibinfo{pages}{107491}.
\newblock \DOIprefix\doi{10.1016/j.compchemeng.2021.107491}.
\bibitem[{Lund et~al.(2018)Lund, Falk, Borup, Madsen and Steen~Mikkelsen}]{lund2018Model}
\bibinfo{author}{Lund, N.S.V.}, \bibinfo{author}{Falk, A.K.V.}, \bibinfo{author}{Borup, M.}, \bibinfo{author}{Madsen, H.}, \bibinfo{author}{Steen~Mikkelsen, P.}, \bibinfo{year}{2018}.
\newblock \bibinfo{title}{Model predictive control of urban drainage systems: A review and perspective towards smart real-time water management}.
\newblock \bibinfo{journal}{Crit. Rev. Environ. Sci. Technol.} \bibinfo{volume}{48}, \bibinfo{pages}{279--339}.
\newblock \DOIprefix\doi{10.1080/10643389.2018.1455484}.
\bibitem[{Maestre et~al.(2012)Maestre, Raso, {van Overloop} and {de Schutter}}]{maestre2012Distributed}
\bibinfo{author}{Maestre, J.M.}, \bibinfo{author}{Raso, L.}, \bibinfo{author}{{van Overloop}, P.J.}, \bibinfo{author}{{de Schutter}, B.}, \bibinfo{year}{2012}.
\newblock \bibinfo{title}{Distributed tree-based model predictive control on an open water system}, in: \bibinfo{booktitle}{Am. {{Control Conf}}.}, \bibinfo{address}{Montreal, QC}. pp. \bibinfo{pages}{1985--1990}.
\newblock \DOIprefix\doi{10.1109/ACC.2012.6314903}.
\bibitem[{Morari and Lee(1999)}]{morari1999Model}
\bibinfo{author}{Morari, M.}, \bibinfo{author}{Lee, H.J.}, \bibinfo{year}{1999}.
\newblock \bibinfo{title}{Model predictive control: past, present and future}.
\newblock \bibinfo{journal}{Comput. Chem. Eng.} \bibinfo{volume}{23}, \bibinfo{pages}{667--682}.
\newblock \DOIprefix\doi{10.1016/S0098-1354(98)00301-9}.
\bibitem[{Negenborn et~al.(2009)Negenborn, van Overloop, Keviczky and Schutter}]{negenborn2009Distributed}
\bibinfo{author}{Negenborn, R.R.}, \bibinfo{author}{van Overloop, P.J.}, \bibinfo{author}{Keviczky, T.}, \bibinfo{author}{Schutter, B.D.}, \bibinfo{year}{2009}.
\newblock \bibinfo{title}{Distributed model predictive control of irrigation canals}.
\newblock \bibinfo{journal}{Networks Heterogen. Media} \bibinfo{volume}{4}, \bibinfo{pages}{359--380}.
\newblock \DOIprefix\doi{10.3934/nhm.2009.4.359}.
\bibitem[{Perelman and Ostfeld(2025)}]{perelman2025Data}
\bibinfo{author}{Perelman, G.}, \bibinfo{author}{Ostfeld, A.}, \bibinfo{year}{2025}.
\newblock \bibinfo{title}{Data enabled predictive control for water distribution systems optimization}.
\newblock \bibinfo{journal}{Water Resour. Res.} \bibinfo{volume}{61}, \bibinfo{pages}{e2024WR039059}.
\newblock \DOIprefix\doi{10.1029/2024WR039059}.
\bibitem[{Potter et~al.(2011)Potter, Wiggert and Ramadan}]{potter2011mechanics}
\bibinfo{author}{Potter, M.}, \bibinfo{author}{Wiggert, D.}, \bibinfo{author}{Ramadan, B.}, \bibinfo{year}{2011}.
\newblock \bibinfo{title}{Mechanics of Fluids}.
\newblock \bibinfo{publisher}{Cengage Learning}.
\bibitem[{Putri et~al.(2024)Putri, Moazeni and Khazaei}]{putri2024Datadriven}
\bibinfo{author}{Putri, S.A.}, \bibinfo{author}{Moazeni, F.}, \bibinfo{author}{Khazaei, J.}, \bibinfo{year}{2024}.
\newblock \bibinfo{title}{Data-driven predictive control strategies of water distribution systems using sparse regression}.
\newblock \bibinfo{journal}{J. Water Process Eng.} \bibinfo{volume}{59}, \bibinfo{pages}{104885}.
\newblock \DOIprefix\doi{10.1016/j.jwpe.2024.104885}.
\bibitem[{Rawlings et~al.(2017)Rawlings, Mayne and Diehl}]{rawlings2017Model}
\bibinfo{author}{Rawlings, J.B.}, \bibinfo{author}{Mayne, D.Q.}, \bibinfo{author}{Diehl, M.}, \bibinfo{year}{2017}.
\newblock \bibinfo{title}{Model Predictive Control: Theory, Computation, and Design}.
\newblock \bibinfo{edition}{2} ed., \bibinfo{publisher}{Nob Hill Publishing}, \bibinfo{address}{Madison, Wisconsin}.
\bibitem[{Ren et~al.(2021)Ren, Niu, Shu, Hancke, Wu, Liu and Xu}]{ren2021Enabling}
\bibinfo{author}{Ren, T.}, \bibinfo{author}{Niu, J.}, \bibinfo{author}{Shu, L.}, \bibinfo{author}{Hancke, G.P.}, \bibinfo{author}{Wu, J.}, \bibinfo{author}{Liu, X.}, \bibinfo{author}{Xu, M.}, \bibinfo{year}{2021}.
\newblock \bibinfo{title}{Enabling efficient model-free control of large-scale canals by exploiting domain knowledge}.
\newblock \bibinfo{journal}{IEEE Trans. Ind. Electron.} \bibinfo{volume}{68}, \bibinfo{pages}{8730--8742}.
\newblock \DOIprefix\doi{10.1109/TIE.2020.3013778}.
\bibitem[{Rentmeesters et~al.(1996)Rentmeesters, Tsai and Lin}]{rentmeesters1996Theory}
\bibinfo{author}{Rentmeesters, M.}, \bibinfo{author}{Tsai, W.}, \bibinfo{author}{Lin, K.J.}, \bibinfo{year}{1996}.
\newblock \bibinfo{title}{A theory of lexicographic multi-criteria optimization}, in: \bibinfo{booktitle}{{{IEEE Int}}. {{Conf}}. {{Eng}}. {{Complex Comput}}. {{Syst}}.}, \bibinfo{address}{Montreal, QC}. pp. \bibinfo{pages}{76--79}.
\newblock \DOIprefix\doi{10.1109/ICECCS.1996.558386}.
\bibitem[{{Rivas-Perez} et~al.(2014){Rivas-Perez}, {Feliu-Batlle}, {Castillo-Garcia} and {Linares-Saez}}]{rivas-perez2014Mathematical}
\bibinfo{author}{{Rivas-Perez}, R.}, \bibinfo{author}{{Feliu-Batlle}, V.}, \bibinfo{author}{{Castillo-Garcia}, F.}, \bibinfo{author}{{Linares-Saez}, A.}, \bibinfo{year}{2014}.
\newblock \bibinfo{title}{Mathematical model for robust control of an irrigation main canal pool}.
\newblock \bibinfo{journal}{Environ. Modell. Software} \bibinfo{volume}{51}, \bibinfo{pages}{207--220}.
\newblock \DOIprefix\doi{10.1016/j.envsoft.2013.10.002}.
\bibitem[{Sch{\"u}tze et~al.(2018)Sch{\"u}tze, Lange, Pabst and Haas}]{schutze2018Astlingen}
\bibinfo{author}{Sch{\"u}tze, M.}, \bibinfo{author}{Lange, M.}, \bibinfo{author}{Pabst, M.}, \bibinfo{author}{Haas, U.}, \bibinfo{year}{2018}.
\newblock \bibinfo{title}{Astlingen -- a benchmark for real time control ({{RTC}})}.
\newblock \bibinfo{journal}{Water Sci. Technol.} \bibinfo{volume}{2017}, \bibinfo{pages}{552--560}.
\newblock \DOIprefix\doi{10.2166/wst.2018.172}.
\bibitem[{Shang et~al.(2024)Shang, Cort{\'e}s and Zheng}]{shang2024Willems}
\bibinfo{author}{Shang, X.}, \bibinfo{author}{Cort{\'e}s, J.}, \bibinfo{author}{Zheng, Y.}, \bibinfo{year}{2024}.
\newblock \bibinfo{title}{Willems' fundamental lemma for nonlinear systems with {{Koopman}} linear embedding}.
\newblock \bibinfo{journal}{IEEE Control Syst. Lett.} \bibinfo{volume}{8}, \bibinfo{pages}{3135--3140}.
\newblock \DOIprefix\doi{10.1109/LCSYS.2024.3522594}.
\bibitem[{Snoek et~al.(2015)Snoek, Rippel, Swersky, Kiros, Satish, Sundaram, Patwary, Prabhat and Adams}]{snoek2015Scalable}
\bibinfo{author}{Snoek, J.}, \bibinfo{author}{Rippel, O.}, \bibinfo{author}{Swersky, K.}, \bibinfo{author}{Kiros, R.}, \bibinfo{author}{Satish, N.}, \bibinfo{author}{Sundaram, N.}, \bibinfo{author}{Patwary, M.}, \bibinfo{author}{Prabhat, M.}, \bibinfo{author}{Adams, R.}, \bibinfo{year}{2015}.
\newblock \bibinfo{title}{Scalable {{Bayesian}} optimization using deep neural networks}, in: \bibinfo{booktitle}{Int. {{Conf}}. {{Mach}}. {{Learn}}.}, \bibinfo{address}{Lille, France}. pp. \bibinfo{pages}{2171--2180}.
\bibitem[{Sorourifar et~al.(2021)Sorourifar, Makrygirgos, Mesbah and Paulson}]{sorourifar2021Datadriven}
\bibinfo{author}{Sorourifar, F.}, \bibinfo{author}{Makrygirgos, G.}, \bibinfo{author}{Mesbah, A.}, \bibinfo{author}{Paulson, J.A.}, \bibinfo{year}{2021}.
\newblock \bibinfo{title}{A data-driven automatic tuning method for {{MPC}} under uncertainty using constrained {{Bayesian}} optimization}.
\newblock \bibinfo{journal}{IFAC-Pap.} \bibinfo{volume}{54}, \bibinfo{pages}{243--250}.
\newblock \DOIprefix\doi{10.1016/j.ifacol.2021.08.249}.
\bibitem[{Stelling and Duinmeijer(2003)}]{stelling2003Staggered}
\bibinfo{author}{Stelling, G.S.}, \bibinfo{author}{Duinmeijer, S.P.A.}, \bibinfo{year}{2003}.
\newblock \bibinfo{title}{A staggered conservative scheme for every froude number in rapidly varied shallow water flows}.
\newblock \bibinfo{journal}{Int. J. Numer. Methods Fluids} \bibinfo{volume}{43}, \bibinfo{pages}{1329--1354}.
\newblock \DOIprefix\doi{10.1002/fld.537}.
\bibitem[{Sutton and Barto(2018)}]{sutton2018Reinforcement}
\bibinfo{author}{Sutton, R.S.}, \bibinfo{author}{Barto, A.G.}, \bibinfo{year}{2018}.
\newblock \bibinfo{title}{Reinforcement Learning: An Introduction}.
\newblock \bibinfo{publisher}{A Bradford Book}, \bibinfo{address}{Cambridge, MA, USA}.
\bibitem[{Ulanicki et~al.(2008)Ulanicki, Kahler and Coulbeck}]{ulanicki2008Modeling}
\bibinfo{author}{Ulanicki, B.}, \bibinfo{author}{Kahler, J.}, \bibinfo{author}{Coulbeck, B.}, \bibinfo{year}{2008}.
\newblock \bibinfo{title}{Modeling the efficiency and power characteristics of a pump group}.
\newblock \bibinfo{journal}{J. Water Resour. Plann. Manage.} \bibinfo{volume}{134}, \bibinfo{pages}{88--93}.
\newblock \DOIprefix\doi{10.1061/(ASCE)0733-9496(2008)134:1(88)}.
\bibitem[{{van der Heijden} et~al.(2022){van der Heijden}, Lugt, {van Nooijen}, Palensky and Abraham}]{vanderheijden2022Multimarket}
\bibinfo{author}{{van der Heijden}, T.}, \bibinfo{author}{Lugt, D.}, \bibinfo{author}{{van Nooijen}, R.}, \bibinfo{author}{Palensky, P.}, \bibinfo{author}{Abraham, E.}, \bibinfo{year}{2022}.
\newblock \bibinfo{title}{Multi-market demand response from pump-controlled open canal systems: An economic {{MPC}} approach to pump-scheduling}.
\newblock \bibinfo{journal}{J. Hydroinf.} \bibinfo{volume}{24}, \bibinfo{pages}{838--855}.
\newblock \DOIprefix\doi{10.2166/hydro.2022.018}.
\bibitem[{{van Overloop} et~al.(2014){van Overloop}, Horv{\'a}th and Ekin~Aydin}]{vanoverloop2014Model}
\bibinfo{author}{{van Overloop}, P.J.}, \bibinfo{author}{Horv{\'a}th, K.}, \bibinfo{author}{Ekin~Aydin, B.}, \bibinfo{year}{2014}.
\newblock \bibinfo{title}{Model predictive control based on an integrator resonance model applied to an open water channel}.
\newblock \bibinfo{journal}{Control Eng. Pract.} \bibinfo{volume}{27}, \bibinfo{pages}{54--60}.
\newblock \DOIprefix\doi{10.1016/j.conengprac.2014.03.001}.
\bibitem[{{van Overloop} et~al.(2008){van Overloop}, Weijs and Dijkstra}]{vanoverloop2008Multiple}
\bibinfo{author}{{van Overloop}, P.J.}, \bibinfo{author}{Weijs, S.}, \bibinfo{author}{Dijkstra, S.}, \bibinfo{year}{2008}.
\newblock \bibinfo{title}{Multiple model predictive control on a drainage canal system}.
\newblock \bibinfo{journal}{Control Eng. Pract.} \bibinfo{volume}{16}, \bibinfo{pages}{531--540}.
\newblock \DOIprefix\doi{10.1016/j.conengprac.2007.06.002}.
\bibitem[{Wang et~al.(2023)Wang, Jin, Schmitt and Olhofer}]{wang2023Recent}
\bibinfo{author}{Wang, X.}, \bibinfo{author}{Jin, Y.}, \bibinfo{author}{Schmitt, S.}, \bibinfo{author}{Olhofer, M.}, \bibinfo{year}{2023}.
\newblock \bibinfo{title}{Recent advances in {{Bayesian}} optimization}.
\newblock \bibinfo{journal}{ACM Comput. Surv.} \bibinfo{volume}{55}, \bibinfo{pages}{287:1--287:36}.
\newblock \DOIprefix\doi{10.1145/3582078}.
\bibitem[{Willems et~al.(2005)Willems, Rapisarda, Markovsky and De~Moor}]{willems2005Note}
\bibinfo{author}{Willems, J.C.}, \bibinfo{author}{Rapisarda, P.}, \bibinfo{author}{Markovsky, I.}, \bibinfo{author}{De~Moor, B.L.M.}, \bibinfo{year}{2005}.
\newblock \bibinfo{title}{A note on persistency of excitation}.
\newblock \bibinfo{journal}{Syst. Control Lett.} \bibinfo{volume}{54}, \bibinfo{pages}{325--329}.
\newblock \DOIprefix\doi{10.1016/j.sysconle.2004.09.003}.
\bibitem[{Williams and Rasmussen(2006)}]{williams2006gaussian}
\bibinfo{author}{Williams, C.K.}, \bibinfo{author}{Rasmussen, C.E.}, \bibinfo{year}{2006}.
\newblock \bibinfo{title}{Gaussian Processes for Machine Learning}. volume~\bibinfo{volume}{2}.
\newblock \bibinfo{publisher}{MIT press Cambridge, MA}.
\bibitem[{Xiong et~al.(2025)Xiong, Yuan, Miao, Wang, Cortes and Papachristodoulou}]{xiong2025Dataenabled}
\bibinfo{author}{Xiong, Z.}, \bibinfo{author}{Yuan, Z.}, \bibinfo{author}{Miao, K.}, \bibinfo{author}{Wang, H.}, \bibinfo{author}{Cortes, J.}, \bibinfo{author}{Papachristodoulou, A.}, \bibinfo{year}{2025}.
\newblock \bibinfo{title}{Data-enabled predictive control for nonlinear systems based on a {{Koopman}} bilinear realization}.
\newblock \DOIprefix\doi{10.48550/arXiv.2505.03346}, \href{http://arxiv.org/abs/2505.03346}{{\tt arXiv:2505.03346}}.
\bibitem[{Yan et~al.(2025)Yan, Zhang, Zhang, Li and Yin}]{yan2025Economic}
\bibinfo{author}{Yan, M.}, \bibinfo{author}{Zhang, X.}, \bibinfo{author}{Zhang, K.}, \bibinfo{author}{Li, Z.}, \bibinfo{author}{Yin, X.}, \bibinfo{year}{2025}.
\newblock \bibinfo{title}{Economic data-enabled predictive control using machine learning}.
\newblock \bibinfo{journal}{IFAC-Pap.} \bibinfo{volume}{59}, \bibinfo{pages}{25--30}.
\newblock \DOIprefix\doi{10.1016/j.ifacol.2025.07.116}.
\bibitem[{Zeng et~al.(2025)Zeng, Cen, Hou, Xie and Chen}]{zeng2025Physicsinformed}
\bibinfo{author}{Zeng, N.}, \bibinfo{author}{Cen, L.}, \bibinfo{author}{Hou, W.}, \bibinfo{author}{Xie, Y.}, \bibinfo{author}{Chen, X.}, \bibinfo{year}{2025}.
\newblock \bibinfo{title}{Physics-informed koopman model predictive control of open canal systems}.
\newblock \bibinfo{journal}{J. Ind. Inf. Integr.} \bibinfo{volume}{46}, \bibinfo{pages}{100845}.
\newblock \DOIprefix\doi{10.1016/j.jii.2025.100845}.
\bibitem[{Zhang et~al.(2023)Zhang, Zheng, Shang and Li}]{zhang2023Dimension}
\bibinfo{author}{Zhang, K.}, \bibinfo{author}{Zheng, Y.}, \bibinfo{author}{Shang, C.}, \bibinfo{author}{Li, Z.}, \bibinfo{year}{2023}.
\newblock \bibinfo{title}{Dimension reduction for efficient data-enabled predictive control}.
\newblock \bibinfo{journal}{IEEE Control Syst. Lett.} \bibinfo{volume}{7}, \bibinfo{pages}{3277--3282}.
\newblock \DOIprefix\doi{10.1109/LCSYS.2023.3322965}.
\bibitem[{Zhang et~al.(2025)Zhang, Zhang, Li and Yin}]{zhang2025Deep}
\bibinfo{author}{Zhang, X.}, \bibinfo{author}{Zhang, K.}, \bibinfo{author}{Li, Z.}, \bibinfo{author}{Yin, X.}, \bibinfo{year}{2025}.
\newblock \bibinfo{title}{Deep {{DeePC}}: Data-enabled predictive control with low or no online optimization using deep learning}.
\newblock \bibinfo{journal}{AIChE J.} \bibinfo{volume}{71}, \bibinfo{pages}{e18644}.
\newblock \DOIprefix\doi{10.1002/aic.18644}.

\end{thebibliography}

\end{document}